\documentclass[a4paper,11pt]{article}
\usepackage[utf8x]{inputenc}
\usepackage{amsfonts}
\usepackage{amsmath}
\usepackage{amssymb}
\usepackage{booktabs}
\usepackage{multirow}
\usepackage{graphicx} 
\usepackage{longtable}
\usepackage[]{threeparttable}

\usepackage{times}
\usepackage{bm}
\usepackage{natbib}
\usepackage[plain,noend]{algorithm2e}

\def\P{{\rm P}}

\def\Bka{{\it Biometrika}}

\def\diag{{\rm diag}}
\def\T{{ \mathrm{\scriptscriptstyle T} }}
\newcommand{\reali}{{\rm I}\negthinspace {\rm R}}

\newcommand\cites[1]{\citeauthor{#1}'s\ (\citeyear{#1})}
\newcounter{example}
\newenvironment{example}[1][]{\refstepcounter{example}\par\medskip\noindent%
   \textbf{Example~\theexample. #1} \rmfamily}{\medskip}
\title{Median bias reduction of maximum likelihood estimates}
\author{E. C. KENNE PAGUI, A. SALVAN and N. SARTORI \\
\small Department of Statistical Sciences, University of Padova \\
\small kenne@stat.unipd.it, salvan@stat.unipd.it,  sartori@stat.unipd.it
}
\begin{document}

\maketitle

\begin{abstract}
\noindent
For regular parametric problems, we show how median centering of the maximum likelihood estimate can be achieved by a simple modification of the score equation. For a scalar parameter of interest, the estimator is equivariant under interest respecting parameterizations and third-order median unbiased. With a vector parameter of interest, componentwise equivariance and third-order median centering are obtained. Like  Firth's (1993, \Bka) implicit method for bias reduction, the new method does not require finiteness of the maximum likelihood estimate and is effective in preventing infinite estimates.
%, although with a reduced shrinkage effect.  
Simulation results for continuous and discrete models, including binary and beta regression, confirm that the method succeeds in achieving componentwise median centering and  in solving the infinite estimate problem, while keeping comparable dispersion and the same approximate distribution as its main competitors.   
\end{abstract}

\noindent
 \emph{Some key words:} Binary regression;   Infinite estimate; Modified score; Parameterization invariance;  Separation problem; Skew normal; Tensor.

%%%%%%%%%%%%%%%%%%%%%%%%%%%%%%
\section{Introduction}
%%%%%I%%%%%%%%%%%%%%%%%%%%%%%%%

In regular parametric estimation problems, both the maximum likelihood estimator and the score estimating function have an  asymptotic symmetric distribution centered at the true parameter value and at zero, respectively. However, the asymptotic behaviour may poorly reflect exact sampling distributions with small or moderate sample information, sparse data or complex models. Several proposals have been developed  to correct the estimate or the estimating function.   

Most available methods are aimed at approximate bias adjustment, either of the maximum likelihood estimator or of the profile score function when nuisance parameters are present.  
We refer to \citet{kosmi2014} for a review of bias reduction   for the maximum likelihood estimator and to \citet{mccu1990}, \citet{stern1997}   and subsequent literature for bias correction of the profile score.

In the absence of nuisance parameters, the score function is exactly unbiased and therefore no correction appears to be necessary.  A change of parameterization does not affect this property  and the solution of the score equation, namely the maximum likelihood estimator, behaves equivariantly under reparameterizations. On the other hand, bias correction of the maximum likelihood estimator is tied to a specific parameterization.  

Lack of equivariance also affects  the so-called implicit bias reduction methods  \citep{kosmi2014}  that  achieve first-oder bias correction through a modification of the score equation, following \citet{firth1993}.  This lack of coherence is highlighted e.g.\ in \citet{kosmi2014}, but somehow overwhelmed by advantages in applications, possibly with a careful choice of the working parameterization \citep[\S~4.2, Remark 3]{kosmi2010}. 
Indeed, one major advantage of the approach in \citet{firth1993} and \citet{kosmi2009} is that the modified estimating equation does not depend explicitly on the maximum likelihood estimate. The modified score equation has been found to overcome infinite estimate problems that may arise with positive probability mainly, but not only, in models for discrete or categorical data.  

Considering first a scalar parameter of interest, we propose a new median modification of the score, or profile score, equation whose solution respects equivariance under monotone reparameterizations.  Like \cites{firth1993} implicit method, this proposal does not rely on finiteness of the maximum likelihood estimate and is effective in preventing infinite estimates. The modification is obtained by considering the median, instead of the mean, as a centering index  for the score and defining a new estimating function by subtracting from the score its approximate median.

Provided that the modified score equation has a unique solution, 
median centering of the score function implies  median centering of the corresponding estimator.  Therefore, the resulting estimator is approximately  median unbiased \citep[see e.g.][]{read1985}, that is 
%it underestimates and overestimates the true parameter value with approximately equal probability, so that 
the true parameter value is approximately a median of the distribution of the estimator.  In some instances exact median unbiased estimates can be obtained \citep[see][]{hirji1989}. Outside exactness cases, available approximations for median unbiased estimates are based on higher-order likelihood asymptotics. Approximations based on the modified signed likelihood ratio \citep{barn1986} have been developed in \citet{pace1999}, \citet{gium2002},   \citet{bieh2015}. They rely, however, on finiteness of the maximum likelihood estimate.   
 Third-order median unbiasedness of the new estimator is seen to hold in the continuous case and extensive numerical evidence indicates remarkable median centering also in the discrete case.
 
We show  how the method can be extended to a vector parameter by simultaneously solving median bias corrected score equations for all parameter components. This leads to componentwise third-order median unbiasedness and parameterization equivariance. 
 
Examples and simulation results in a number of models, including binary and beta regression, indicate that the new estimator provides a notable improvement over the maximum likelihood estimator and solves the infinite  estimates problem, both for a scalar and for a vector parameter.

%%The construction is conceptually straightforward for a scalar parameter of interest, even in the presence of nuisance parameters  if the   profile score is considered. 
%Provided that the modified score equation has a unique solution, 
%%Under monotonicity assumptions, 
%median centering of the score function implies  median centering of the corresponding estimator.  Therefore, the resulting estimator is third-order  median unbiased \citep[see e.g.][]{read1985}, that is it underestimates and overestimates the true parameter value with approximately equal probability, so that the true parameter value is approximately a median of the distribution of the estimator.  In some instances exact median unbiased estimates can be obtained \citep[see][]{hirji1989}.
%%, median unbiased estimates may represent an alternative to \cites{firth1993} bias preventing method (see for instance the {\tt proc logistic} in SAS). 
%Outside exactness cases, available approximations for median unbiased estimates are based on higher-order likelihood asymptotics. Approximations based on the modified signed likelihood ratio \citep{barn1986} have been developed in \citet{pace1999}, \citet{gium2002},   \citet{bieh2015}. They rely, however, on finiteness of the maximum likelihood estimate.   
%%, and produces a shrinkage ... less than Firth (1993).
% 
% 

 %%%%%%%%%%%%%%%%%%%%%%%%%%%%%%
\section{Median modified score  for a scalar parameter of interest}
%%%%%%%%%%%%%%%%%%%%%%%%%%%%%%

%While the score has mean exactly equal to zero, only the leading term of its median is equal to zero. In the following we define a median modified score equation having median zero with error of order $O(n^{-1})$. 
%
%For a scalar parameter of interest, we introduce the median modified score with no nuisance parameters in Section \ref{no_nuisance}.  In Section \ref{nuisance} we extend the modification to the profile score when nuisance parameters are present. The latter result is a necessary step for the development of a median modified score for all components of a vector parameter in Section \ref{vector}.      

 %%%%%%%%%%%%%%%%%%%%%%%%%%%%%%
\subsection{No nuisance parameters}\label{no_nuisance}
%%%%%%%%%%%%%%%%%%%%%%%%%%%%%%
%Consider first the case of a scalar $\theta$ and no additional nuisance parameters.  
For data $y$, consider a regular model with probability mass function $p_{_Y} (y; \theta)$, $\theta\in\Theta\subseteq \reali$. 
Let $\ell(\theta)$ be the corresponding log likelihood and  $U=U(\theta)=\partial \ell(\theta)/\partial  \theta$, the score function. The maximum likelihood estimator $\hat\theta$ is   solution of $U(\theta)=0$. We assume that Fisher information, $i(\theta)$,  and the third-order cumulant of $U(\theta)$ are finite and of order $O(n)$, where $n$ is the sample size or, more generally, an index  of information in the data.  

Using Cornish-Fisher expansion \citep[see e.g.][\S~10.6]{pace:salv:1997}, the following asymptotic expansion  holds for the median under $\theta$, $M_\theta(\cdot)$,  of the score  in the continuous case
\begin{equation*}\label{Cornish}
M_{\theta}\left\{U(\theta)\right\}= - \nu_{\theta,\theta,\theta}/\{6\, i(\theta)\} + O(n^{-1})\,,
\end{equation*}
with $\nu_{\theta,\theta,\theta}=\nu_{\theta,\theta,\theta}(\theta)= E_{\theta}\{U(\theta)^3 \}$.
% and $i(\theta)$ is Fisher information.
A modified score equation can thus be defined by equating $U(\theta)$ to the leading term of its median. This suggests defining the median modified score
\begin{equation}\label{modscore}
\tilde{U}(\theta)= U(\theta) + \nu_{\theta,\theta,\theta}/\{6\, i(\theta)\} \,,
\end{equation}
where the modification term $\nu_{\theta,\theta,\theta}/\{6\, i(\theta)\}$ is of order $O(1)$. 
Let  $\tilde\theta$  be the estimator defined as solution of $\tilde{U}(\theta)=0$. 

For $\tilde{U}(\theta)$, we have $M_{\theta} \{\tilde{U}(\theta) \}=O(n^{-1})$ and it is shown in the Appendix that $\tilde{U}(\theta)$ is  third-order median unbiased, i.e.\  
\begin{equation}\label{mu_modscore}
\P_{\theta}\{\tilde{U}(\theta)\leq 0 \}= 1/2 + O(n^{-3/2})\,.
\end{equation} 
If $\tilde\theta$ is the unique solution of $\tilde{U}(\theta)=0$, 
%***or, at least, the probability of multiple roots is  $O(n^{-3/2})$,*** 
 %is monotone decreasing in $\theta$, 
the events
$
\tilde{U}(\theta)\leq 0$ and $\tilde\theta \leq \theta 
$
are equivalent so that  $\tilde\theta$ will be  third-order median unbiased, i.e.
\begin{equation}\label{mu}
\P_{\theta}\{\tilde\theta\leq \theta \}= 1/2+ O(n^{-3/2})\,.
\end{equation}
Like $\hat\theta$, 
%and $\hat\theta^*$, 
also $\tilde\theta$ is asymptotically $N\{\theta, i(\theta)^{-1}\}$, so that  Wald-type confidence intervals only differ in location. Score-type confidence intervals can also be used, based on the asymptotic $N\{0, i(\theta)\}$ distribution of $\tilde{U}(\theta)$.

If $\omega(\theta)$ is a smooth reparameterization with inverse $\theta(\omega)$, ingredients of the modification term in  (\ref{modscore}) in the new parameterization are 
$
\nu^{\Omega}_{\omega,\omega,\omega}=\nu_{\theta,\theta,\theta}\{\theta(\omega)\} \{\theta'(\omega)\}^3
$
and 
$
i^{\Omega}(\omega)=i\{\theta(\omega)\} \{\theta'(\omega)\}^2
$, where $\theta'(\omega)=d\theta(\omega)/d\omega$. Hence, like $U(\theta)$, the modified score $\tilde{U}(\theta)$ transforms as a covariant tensor of order one,  namely the modified score in the $\omega$ parameterization is 
$\tilde{U}\{\theta(\omega)\} \theta'(\omega)$. 
Therefore, $\tilde{\theta}$ behaves equivariantly as does $\hat\theta$, and $\tilde\omega=\omega(\tilde\theta)$ is also third-order median unbiased.   

%On the other hand, 
\cites{firth1993}  method gives an estimator $\hat\theta^*$ with bias of order $O(n^{-2})$ in a chosen parameterization. For a scalar parameter, the corresponding modified score is 
\begin{equation}\label{firth1}
U^*(\theta)=U(\theta)+(\nu_{\theta,\theta,\theta}+\nu_{\theta,\theta\theta})/\{2i(\theta)\}\,,
\end{equation}
where $\nu_{\theta,\theta\theta}=E_{\theta}\left\{U(\theta) U_{\theta\theta}(\theta)\right\}$, with
$U_{\theta\theta}(\theta)=\partial^2 \ell(\theta)/\partial\theta^2$. As shown by \citet[\S~3.4]{kosmi2010} in the vector parameter case, $U^*(\theta)$ does not transform as a covariant tensor of order one under reparameterizations. This is because, while  $i(\theta)$ behaves tensorially, the same is not true for the term $\nu_{\theta,\theta,\theta}+\nu_{\theta,\theta\theta}$. Therefore, as is natural, first-order bias correction only operates in the reference parameterization. 
A suggestion in  \citet[\S~4.2, Remark 3]{kosmi2010} is to obtain the correction in a parameterization where the distribution of the maximum likelihood estimator is closer to normality, such as the logit for probability parameters, and then translate the result in the parameterization of interest.

The argument leading to (\ref{mu_modscore}) and (\ref{mu}) only holds in the continuous case. Indeed, 
in the discrete case, the Cornish-Fisher expansion involves also oscillatory terms \citep[see e.g.][formula (A.1)]{cai2009}.  These terms will be ignored in the following and the same adjustment will be employed both in the continuous and in the discrete case.  Empirical results in the paper  show a gain in median unbiasedness using (\ref{modscore}) in place of $U(\theta)$ also in the discrete case.  The effect of omitting the oscillatory terms in a simple logistic regression is illustrated in detail in the Supplementary Material, showing that $\tilde{\theta}$ is uniformly closer to the exact median unbiased estimator than $\hat\theta$. Moreover, as the number of points in the support of the sufficient statistic increases, $\tilde{\theta}$ gets much closer to the exact median unbiased estimator than $\hat\theta$.

For a  one parameter exponential family with canonical parameter $\theta$, i.e.\ with
\begin{equation}\label{eq1}
p_{_Y}(y; \theta)=\exp\{\theta t(y)-K(\theta)\} h(y)\,,
\end{equation}
the median modified score function has the form
  \begin{equation*}
 \begin{split}
\tilde U(\theta)= U(\theta) +K_{\theta\theta\theta}/\{6\,K_{\theta\theta}\}\,,
\end{split}
 \end{equation*}
 where $K_{\theta\theta\theta}=\partial^3 K(\theta)/\partial\theta^3 $ and $K_{\theta\theta}=\partial^2 K(\theta)/\partial\theta^2 =i(\theta)$.   
In this parameterization,  $\tilde U(\theta)$  can be seen as the score of the penalized log likelihood 
$%  \begin{equation*}
 \tilde\ell(\theta)=\ell(\theta)+\{\log i(\theta)\}/6.
% \tilde\ell(\theta)=\ell(\theta)+(\log K_{\theta\theta})/6=\ell(\theta)+\{\log i(\theta)\}/6\,.
$ % \end{equation*}
On the other hand, \cites{firth1993} modified score takes the form 
\begin{equation}\label{firth_expfam1}
 \begin{split}
U^*(\theta)= U(\theta) + K_{\theta\theta\theta}/\{2\,K_{\theta\theta}\}\,.
\end{split}
 \end{equation}
The effect of the median modification is thus to penalize the likelihood by $i(\theta)^{1/6}$, while (\ref{firth_expfam1}) implies a Jeffreys prior penalization. 
%implied by  This reflects on the shrinkage effect, which is smaller for $\tilde\theta$ than for $\hat\theta^*$. 

Under model (\ref{eq1}),  $U(\theta)=t(y)-E_{\theta}(t(Y))$, hence, if $K(\theta)=O(n)$,  the estimating equation $\tilde{U}(\theta)=0$ provides, in the continuous case, an approximate version of the optimal median unbiased estimator for monotone likelihood ratio families, calculated as the value $\tilde\theta^e$ of $\theta$ such that  
  $\P_{\theta}(T\leq t) =1/2$ 
 \citep[\S~3.5]{lehm2005}.  
Use of  $\tilde{U}(\theta)=0$ amounts to replace the exact   $\P_{\theta}(T\leq t) $ with its Edgeworth expansion up to terms of order $O(n^{-1})$. It is straightforward to see that $\tilde\theta-\tilde\theta^e=O_p(n^{-2})$.

In general, a regular model has locally a monotone likelihood ratio with respect to the score function \citep[\S~4.8.i]{cox74}. 
%, that is, denoting by $\theta_0$ the true parameter value, and with $\delta\in \reali$, 
%$$
%\log \frac{p_{_Y}(y;\theta_0+\delta)}{p_{_Y}(y;\theta_0)}=\delta U(\theta_0)+o(\delta)\,.
%$$
As a consequence, optimality of $\tilde\theta$ as defined e.g.\ in \citet[formula (3.58)]{pace:salv:1997} will hold locally in a neighbourhood of $\theta_0$. 

\begin{example}\label{ex-normal} 
% va messo un \mu_0 come argom o deponente delle varie stime??
%estimator o estimate??
%meglio theta al posto di \psi??
Normal distribution with known mean.
Let $y_1,\ldots,y_n$ be a random sample from $N(\mu, \psi)$, with known $\mu$.   Quantities for computing  (\ref{modscore}) and   (\ref{firth1}) are given in \citet[][\S~4.2]{firth1993}. In particular, the adjustment in (\ref{firth1}) is equal to zero, so that   $\hat\psi= \hat\psi^{*}=  s(\mu)/n$, with $s(\mu)=\sum_{i=1}^n (y_i-\mu)^2$, is exactly unbiased.    The median modified score (\ref{modscore})  is equal to
$-(n-2/3)/(2\sigma^2) + s(\mu)/(2\sigma^2)^2$,  giving $\tilde\psi=
s(\mu)/(n-2/3)$, %that is 
equal to the optimal median unbiased estimator $s(\mu)/\chi^2_{n;0.5}$  plus an error of order $O(n^{-2})$. Consider now the parameterization with the standard deviation  $\omega=\psi^{1/2}$. By equivariance, $\hat\omega=\hat\psi^{1/2}$ and $\tilde\omega=\tilde\psi^{1/2}$, while the bias reduced estimator calculated in the new parameterization is $\hat\omega^*=\{s(\mu)/(n-1/2)\}^{1/2}$.
 \end{example}

\begin{example}\label{ex-SN} 
Skew normal shape parameter.
Let $y_1,\ldots,y_n$ be $n$ independent realizations of a skew normal distribution with shape parameter $\theta\in\reali$ and density
  $p(y;\theta)=2\phi(y)\Phi(\theta y)$, where $\phi$ and $\Phi$ denote the standard normal density and distribution
functions, respectively, and $y\in\reali$. The log likelihood  is
$
\ell(\theta) =  \sum_{i=1}^n \zeta_0(\theta y_i),
$
where $\zeta_0(x)=\log\{2\Phi(x)\}$. With $\zeta_m(x)=\partial^m \zeta_0(x)/ \partial x^m$, $m=1,2,\ldots$,
the score function is 
$
U(\theta)= \sum_{i=1}^n \zeta_1(\theta y_i)y_i.
$ 
Let  $a_{kh}(\theta)=E_{\theta}\{Y^k \zeta_1(\theta Y)^h\}$.
%, with $a_{k1}(\theta)=0$ when $k$ is odd, $a_{kh}(\theta)\geq 0$ when
%both $k$ and $h$ are even. 
The expected quantities needed to compute the median modified score  (\ref{modscore}) are
 $i(\theta)=  na_{22}(\theta)$ and $\nu_{\theta,\theta,\theta} =na_{33}(\theta)$, 
giving
%$$
%\tilde U(\theta)=U(\theta)+\frac{1}{6}\frac{a_{33}(\theta)}{a_{22}(\theta)}\,.
%$$
$
\tilde U(\theta)=U(\theta)+a_{33}(\theta)/\{6 a_{22}(\theta)\}
$. The modified score (\ref{firth1})  \citep[see][]{sartori2006}  is 
%$$
%U^*(\theta)=U(\theta)-\frac{\theta}{2}\frac{a_{42}(\theta)}{a_{22}(\theta)}\,.
%$$
$
U^*(\theta)=U(\theta)- \theta a_{42}(\theta)/\{2 a_{22}(\theta)\}
$.

The performance of  $\hat \theta$, $\hat\theta^*$ and $\tilde \theta$ has been investigated by Monte Carlo simulations with 5,000 replications. Results are displayed in Table \ref{tab1}.  Estimators are compared in terms of empirical probability of underestimation,  median absolute error, bias,  root mean squared error  and coverage of 95\% Wald-type and score-type confidence intervals. The empirical probability of underestimation is  the  summary of primary interest for $\tilde\theta$,   as the estimator is designed to satisfy (\ref{mu}). A natural associated measure of dispersion is the median absolute error. Estimated bias and root mean squared error are also reported  to enable a fair comparison with $\hat\theta^*$.  While  $\hat\theta^*$ and $\tilde\theta$ are always finite, in some samples the maximum likelihood estimate is infinite.  The simulation frequency of finite maximum likelihood estimates, $\%$($\hat\theta < +\infty$), is reported in the table.  As in \citet[][\S~6.2]{kosmi2009},   estimated bias, root mean squared error and coverage probability of confidence intervals for $\hat\theta$ are conditional upon its finiteness.  Although this favours  
$\hat\theta$, both $\tilde\theta$ and $\hat\theta^*$ are uniformly better.   
Median centering improvement attained by $\tilde\theta$, as measured  by empirical probability of underestimation, is remarkable, both for small and moderate sample sizes.  
On the other hand,   the estimated root mean squared error is much smaller for of $\hat\theta^*$  than for $\tilde\theta$. 
In this case, values of $\tilde\theta$ are intermediate between those of $\hat\theta$ and $\hat\theta^*$. This effect is  illustrated, for the same sample as in Example 1 of 
\citet{sartori2006}, in Fig.\ \ref{fig-skew}. Score-type confidence intervals have overall better coverage than Wald-type intervals, although this effect  is substantial only for $\hat\theta^*$. Indeed, 
the penalization implied by $\hat\theta^*$ is excessive, leading  to poor coverage of Wald-type confidence intervals. Coverage probabilities for maximum likelihood should be judged with caution since samples with infinite estimates are excluded. 

%A further example with the same simple discrete data model as in  \citet{firth1993}   is given in Example 1 in the Supplementary Material. 

\begin{table}\centering
 \begin{threeparttable}
 \def~{\hphantom{0}}
\caption{Simulation results for estimates of the skew normal shape parameter.
%, using maximum likelihood $\hat\theta$, bias reduction $\hat\theta^*$ and median bias reduction $\tilde\theta$. 
For $\hat\theta$,  B, RMSE and coverage are conditional upon finiteness of the estimates}{
\begin{tabular}{cclccccccc}
% \toprule
% \toprule
$ \theta$&$n$&~& PU & MAE& B & RMSE & Wald  & Score & $\%$($\hat\theta < +\infty$) \\ 
% \hline
5& 20&$\hat\theta$ & 36.2   & 2.31        & 1.90 & 8.44         &  94.5& 94.7  & 72.2  \\ 
% & &$\hat\theta_p^*$ & 92.70  & 1.89        & -1.68 & 2.00 & 69.0\\
& &$\hat\theta^*$ & 92.8    & 1.91        & -1.70 & 2.01         & 68.4 & 87.0 & \\ 
& &$\tilde\theta$ & 53.8    & 1.73        & 0.94 & 4.02          & 91.1 & 92.5 &\\ 
& 50&$\hat\theta$ & 41.0    & 1.31        & 1.93 & 8.67          &  96.5 & 95.0 &  96.0  \\ 
%& & $\hat\theta_p^*$ & 67.4 & 1.19    & -0.27  & 1.79          & 86.5\\
& &$\hat\theta^*$ & 67.7    & 1.20        & -0.28 & 1.79         & 86.2 & 90.3&\\ 
& &$\tilde\theta$ & 50.3    & 1.21        & 1.25 & 4.82        & 93.9 & 93.4&\\ 
& 100&$\hat\theta$ & 42.7   & 0.86        & 0.82 & 3.64         & 96.9 & 96.1 &99.9  \\  
%& & $\hat\theta_p^*$ & 60.9 & 0.84    & 0.01  & 1.50        & 91.9 \\
& &$\hat\theta^*$ & 60.9    & 0.84        & 0.00 & 1.50          &  91.9 & 92.8&\\ 
 & &$\tilde\theta$ & 49.8   & 0.84        & 0.49 & 2.20         & 95.5 & 95.3&\\ 
10 &  20&$\hat\theta$& 29.7 & $+\infty$    & 2.12 & 20.11      &  90.6 & 93.9 & 49.2   \\ 
% & & $\hat\theta_p^*$ & 99.8 & 6.14        & -5.92 & 6.04        & 20.7\\
& &$\hat\theta^*$ & 99.8    & 6.16        & -5.94 & 6.06        & 20.4 & 83.7&\\ 
 & &$\tilde\theta$ & 73.0   & 3.57        & -1.36 & 5.35        & 83.2 & 91.7&\\ 
& 50&$\hat\theta$ & 36.9    & 3.73        & 5.11 & 30.11         & 95.5   & 95.2  & 80.2  \\ 
%  & & $\hat\theta_p^*$         & 87.2     & 3.24  & -2.59          & 3.55     & 73.7\\
 & &$\hat\theta^*$  & 87.2  & 3.25        & -2.59 & 3.56       & 73.5 & 88.4 &\\ 
& &$\tilde\theta$ & 52.6    & 3.10        & 2.30 & 8.67          & 92.0 & 93.4&\\ 
 & 100&$\hat\theta$ & 40.1  & 2.50        & 3.92 & 15.95       & 96.1 & 95.4& 96.0 \\
%   & & $\hat\theta_p^*$ & 67.8 &  2.28   & -0.50 & 3.54        & 87.0\\
  & &$\hat\theta^*$ & 68.0  & 2.28        & -0.52 & 3.53       & 86.9 & 90.8&\\ 
& &$\tilde\theta$ & 49.6    & 2.32        & 2.57 & 10.01         & 93.9 & 93.9&\\ 
\end{tabular}}
 \label{tab1}
\begin{tablenotes}\footnotesize\item
PU, percentage of underestimation; MAE,   median absolute error; B, bias; RMSE, root mean squared error; Wald and Score, percentage coverage of 95\% Wald-type and score-type confidence intervals.  
%Wald-type confidence intervals (95\%CI). 
%U.S., United States of America; R, respondent.
\end{tablenotes}
\end{threeparttable}
\end{table}

\begin{figure}
%\vspace{-1cm}
\includegraphics[width=0.95\textwidth]{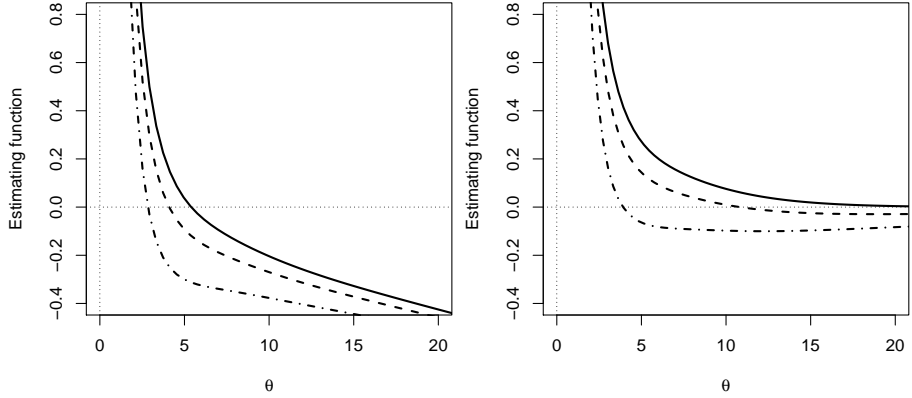} 
%&\figurebox{20pc}{25pc}{}[scores2]
%\includegraphics[]{scores} & 
%\includegraphics[]{scores2} 
%\end{tabular}
%\vspace{-1.cm}
\caption{The left panel shows $U(\theta)$ (solid), $U^*(\theta)$ (dot-dashed) and $\tilde U(\theta)$ (dashed) with data from  \citet[Example 1]{sartori2006}  
with corresponding estimates 5.40,  2.84 and  4.10, respectively. The right panel is relative 
to the same data with a change of sign of the only negative observation, with estimates $+\infty$,  3.92 and 10.82.}
\label{fig-skew}
\end{figure}
\end{example}

%%%%%%%%%%%%%%%%%%%%%%%%%%%%%%%%%%%%%%%%%%%%
\subsection{Presence of nuisance parameters}\label{nuisance}
%%%%%%%%%%%%%%%%%%%%%%%%%%%%%%%%%%%%%%%%%%%%

With $\theta=(\theta_1,\ldots,\theta_p)$, we denote by $U_r=\partial \ell(\theta)/\partial  \theta^r$, $r=1,\ldots,p$, 
the elements of the score vector $U(\theta)$. 
Let $i_{rs}$ be a generic entry of Fisher information, $i(\theta)$, and $i^{rs}$ an entry of its inverse, $r,s,\ldots=1,\ldots,p$.
 Let  $U_{rs}$ and $U_{rst}$  be higher order partial derivatives of $\ell(\theta)$ with respect to elements of $\theta$ with indices $r,s,t $.  
Moreover,  expected values of log likelihood derivatives are  denoted as 
$\nu_{rs}=E_{\theta}(U_{rs})=-i_{rs}$, $\nu_{rst}=E_{\theta}(U_{rst})$, $\nu_{r,st}=E_{\theta}(U_{r}U_{st})$ and 
$\nu_{r,s,t}=E_{\theta}(U_{r}U_{s}U_{t})$.

Let us suppose  now that the parameter is partitioned as  $\theta=(\psi,\lambda)$,  with $\psi$ a scalar parameter of interest. When exact elimination of $\lambda$ by conditioning or by marginalization  is feasible,  arguments in the previous subsection may be applied to the conditional or marginal score for $\psi$. See e.g.\  \citet{hirji1989} for exact conditional median unbiased estimators in logistic regression. In more general situations, or when an expression for the exact solution is not available, we propose a modification of the profile score.    Let us denote by $\ell_{_P}(\psi)=\ell(\psi, \hat\lambda_\psi)$ the profile log likelihood for $\psi$, where $\hat\lambda_{\psi}$ is the maximum likelihood estimate of $\lambda$ for a given value of $\psi$. The profile score is $U_{_P}(\psi)=\partial \ell_{_P}(\psi)/\partial\psi$.  Let us use subscript $\psi$ when referring to $\psi$ and indices $a,b,c,\ldots$ to refer to components of $\lambda$, so that elements of  $U(\theta)$ are $U_\psi=U_\psi(\psi,\lambda)=\partial \ell(\psi,\lambda)/\partial\psi$ and $U_a=U_a(\psi,\lambda)=\partial \ell(\psi,\lambda)/\partial\lambda_a$, $a=1,\ldots, p-1$.
As is well known, $U_{_P}(\psi)=U_\psi(\psi,\hat\lambda_\psi)$ and approximate expressions for the first three cumulants of $U_{_P}(\psi)$ are  
 \begin{equation}\label{kprof}
\begin{split}
\kappa_{1\psi} &= 
-\frac{1}{2}\nu^{ab}\{(\nu_{\psi, ab}-\gamma_{\psi c}\nu_{c,ab})+(\nu_{\psi, a,b}- \gamma_{\psi c}\nu_{a,b,c})\}\\
\kappa_{2\psi} & = \nu_{\psi,\psi}-\gamma_{\psi a}\nu_{\psi, a} \\
\kappa_{3\psi} & = \nu_{\psi,\psi,\psi}-3\gamma_{\psi a}\nu_{\psi,\psi, a}+3\gamma_{\psi a} \gamma_{\psi b}\nu_{\psi, a,b }-\gamma_{\psi a} \gamma_{\psi b} \gamma_{\psi c}\nu_{a,b,c} \,,\end{split}
\end{equation}
 where the error term is of order $O(n^{-1})$ in $\kappa_{1\psi}$ and of order $O(1)$ in $\kappa_{2\psi}$ and $\kappa_{3\psi}$.
 In (\ref{kprof}), Einstein summation convention is used, i.e.\, summation over repeated indices $a,b,\ldots$ is understood. The quantity  
$\nu^{ab}$ is an element of the inverse of the square matrix of order $p-1$ with entries $\nu_{a,b}$, and  $\gamma_{\psi a}=\nu^{ab}\nu_{\psi,b}$ is a regression coefficient of $U_\psi$ on the vector with elements $U_a$, $a=1,\ldots,p-1$. The above expression for $\kappa_{1\psi}$ was obtained in \citet{mccu1990}. Approximations $\kappa_{2\psi}$ and $\kappa_{3\psi}$ are the second and third cumulants of the efficient score for $\psi$, namely $\bar{U}_\psi=U_\psi-\gamma_{\psi a}U_a$, which is the leading  term of the expansion of $U_{_P}(\psi)$. They   are obtained from formulae (7.15) and (7.16) in \citet{bncox89} for cumulants of residuals.
 
In a continuous model, using a  Cornish-Fisher expansion, the median of the standardized profile score $(U_{_P}(\psi)-\kappa_{1\psi})/\sqrt{\kappa_{2\psi}}$ is equal to $-\kappa_{3\psi}/(6\kappa_{2\psi}^{3/2})+O(n^{-3/2})$. 
 Therefore, the median modified profile score is 
  \begin{equation}\label{modprofscore}
 \tilde U_{_P}(\psi)=U_{_P}(\psi)-\kappa_{1\psi}+\kappa_{3\psi}/\{6\,\kappa_{2\psi}\}
 \end{equation}
 and  has median zero with error of order $O(n^{-1})$. The same argument as in the proof of (\ref{mu_modscore}) shows that $\P_{\theta}(\tilde U_{_P}(\psi)\leq 0)=1/2+O(n^{-3/2})$. 
 Let $\tilde\psi_{_P}$ be the estimator defined as solution of  $ \tilde U_{_P}(\psi)=0$ with $\lambda$  replaced by $\hat\lambda_\psi$. If the resulting estimating equation 
 has a unique solution, 
 % ***or, at least, the probability of multiple roots is  $O(n^{-3/2})$,*** 
 %is monotone decreasing in $\psi$, 
 third-order median unbiasedness of $\tilde\psi_{_P}$ follows. Although this argument only holds in the continuous case, empirical results for binary regression in Examples 4 and 7  show a gain in median unbiasedness using (\ref{modprofscore}) in place of $U_{_P}(\theta)$ also in the discrete case. See the Supplementary Material for a numerical comparison of $\tilde\psi_{_P}$ with the exact conditional median unbiased estimator. The asymptotic distribution of $\tilde\psi_{_P}$ is the same as that of $\hat\psi$, that is $N(\psi, \kappa_{2\psi}^{-1})$. This can be used to construct Wald-type confidence intervals. 
 Score-type confidence intervals can also be used, based on the asymptotic $N(0, \kappa_{2\psi})$ distribution of $\tilde U_{_P}(\psi)$.

 Substituting  $\hat\lambda_\psi$  for $\lambda$ has the drawback of requiring  the solution of $U_a=0$ for fixed $\psi$, $a=1,\ldots,p-1$. Although infinite values of the constrained estimate of $\lambda$ may not be a problem in (\ref{modprofscore}),  joint estimation as described in \S~3 is often preferable.

Parameterization equivariance of $\tilde\psi_{_P}$ holds under interest respecting reparameterizations.
In detail, let $\omega=(\varphi,\chi)$ be a smooth reparameterization  with $\varphi=\varphi(\psi)$ and $\chi=\chi(\psi,\lambda)$ and $\varphi=\varphi(\psi)$ a one-to-one function of $\psi$ with inverse $\psi(\varphi)$. Then, the modified score for $\varphi$ in the new parameterization is $\tilde U_{_P}(\psi(\varphi)) \psi'(\varphi)$, so that $\tilde\varphi_{_P}=\varphi(\tilde\psi_{_P})$. This tensorial behaviour of the modified profile score  follows from the tensorial behaviour of the profile score and of its first-order expectation  \citep[\S~9.5.3]{pace:salv:1997}. In addition,  the efficient score $\bar{U}_\psi$ also transforms tensorially and therefore so does the  ratio $\kappa_{3\psi}/\kappa_{2\psi}$.

If $p_{_Y}(y;\theta)$ is an exponential family of order $p$ with canonical parameter $(\psi,\lambda)$, i.e.\ 
\begin{equation*}\label{eq2}
p_{_Y}(y; \psi,\lambda)=\exp\{\psi t(y)+\lambda^\T s(y)-K(\psi,\lambda)\} h(y)\,,
\end{equation*}
quantities (\ref{kprof}) are simply obtained from derivatives of $K(\psi,\lambda)$. In particular, $\nu^{ab}$ is a generic element of $(\partial^2 K(\psi,\lambda)/\partial\lambda\partial\lambda^\T)^{-1}$, $\nu_{\psi, ab}=\nu_{c,ab}=0$, and all other $\nu$ quantities are the derivatives of $K(\psi,\lambda)$ with respect to components of $(\psi,\lambda)$ appearing as subscripts. Here,  $U_{_P}(\psi)-\kappa_{1\psi}$ is an approximation with error of order $O(n^{-1})$ of the   score for $\psi$ in the conditional  model given $s(y)$ \citep[see e.g.][\S~10.10.2]{pace:salv:1997}. In the continuous case, the estimator from (\ref{modprofscore}) is an approximation  of the optimal conditional median unbiased estimator \citep[\S~5.4]{lehm2005}, solution with respect to $\psi$ of $\P_{\psi}(T\leq t|S=s) =1/2$.
%\begin{equation}\label{mueexpfamcond}
%\begin{split}
%\P_{\psi}(T\leq t|S=s) = \frac{1}{2}\,.
%\end{split}
%\end{equation}
The approximation is obtained by replacing $\P_{\psi}(T\leq t|S=s)$ with its mixed Edgeworth-saddlepoint approximation (\citealp[][\S~7.5]{bncox89}, \citealp{pacesalvan1992})
 up to terms of order $O(n^{-1})$.

In the examples below,   $\tilde\psi_{_P}$ is compared with  $\hat\psi$ and with   $\hat\psi^*$, i.e.\ the $\psi$ component of the bias reduced maximum likelihood estimator $\hat\theta^*$, calculated according to formula (4.1) in \citet{firth1993}. 

\begin{example}\label{ex-normal-two} 
% estimate o estimator??
Normal distribution (cont.).
Consider again the setting of Example \ref{ex-normal} with both $\mu$ and  $\psi$ unknown and let $\psi$ be of interest. 
The maximum likelihood estimator is $ \hat\psi= s(\bar{y})/n$, with $\bar{y}=\sum_{i=1}^n y_i/n$. \citet[][\S~4.2]{firth1993} shows that $\hat\psi^{*}=s(\bar{y})/(n-1)$,  which coincides with the usual  unbiased estimator. Formula (\ref{modprofscore})   gives
$\tilde{U}_{_P}(\psi)= -(n-1-2/3)/(2\psi)+s(\bar{y})/(2\psi^2)$, so that 
$\tilde\psi_{_P}=  s(\bar{y})/(n-1-2/3)$, that is equal to the optimal median unbiased estimator $s(\bar{y})/\chi^2_{n-1;0.5}$ plus an error of order $O(n^{-2})$. In the $(\mu,\omega)$ parameterization, with
$\omega=\psi^{1/2}$, the bias reduced estimator is $\hat\omega^*=\{s(\bar{y})/(n-3/2)\}^{1/2}$.   
 \end{example}

\begin{example}\label{binaryregression} 
 Binary regression.
  Let $y_i$, $i=1,\ldots,n$, be independent realizations of binary random variables with probability $\pi_i=F(\eta_i)$, where $\eta_i= x_i\beta$, $x_i=(x_{i1},\ldots, x_{ip})$ is a row vector of covariates and $F$ is a known cumulative distribution function. We assume that a generic scalar component of   $\beta$ is of interest and treat the remaining components as nuisance parameters. Quantities needed for (\ref{modprofscore})  are given in the Supplementary Material.
%  \begin{eqnarray*}
%U_r &=&\sum_{i=1}^n x_{ir} A_i\{ y_i-F(\eta_i)\}\qquad 
%i_{rs}= \sum_{i=1}^n x_{ir} x_{is} A_i F'(\eta_i)\\
%\nu_{rs,t} &=& \sum_{i=1}^n x_{ir} x_{is}x_{it} B_i F'(\eta_i)\\
%\nu_{r,s,t}&=& \sum_{i=1}^n x_{ir} x_{is}x_{it} A_i^3 F(\eta_i)\{ 1-F(\eta_i)\} \{ 1-2F(\eta_i)\} \,,
%\end{eqnarray*}
%with
%\begin{eqnarray*}
%A_i &=&\frac{F'(\eta_i)}{F(\eta_i)\{ 1-F(\eta_i)\}}\\
%B_i &=&\frac{F''(\eta_i)}{F(\eta_i)\{ 1-F(\eta_i)\}}+\frac{F'(\eta_i)^2\{ 2F(\eta_i)-1\}}{F(\eta_i)^2\{ 1-F(\eta_i)\}^2} \,,
%\end{eqnarray*}
%
%\noindent
%were $F'(\cdot)$ and $F''(\cdot)$ are first and second derivatives of $F(\cdot)$. If $F(\cdot)$ is the logistic cumulative distribution function, $A_i=1$ and $B_i=0$.

As an example, we consider the endometrial cancer grade dataset analyzed, among others, 
% in  \citet[\S~4.1]{hein2002} and 
 in \citet[\S~5.7.1]{agre2015}. The goal of the study  was to evaluate the relationship between the histology of the endometrium of 79 patients and three risk factors:   neovasculation,  pulsatility index of arteria uterina and endometrium height. 
Logistic regression has been fitted with parameter $\beta=(\beta_1, \beta_2, \beta_3, \beta_4)^\T$, where $\beta_1$ is an intercept and the remaining  parameters correspond to  neovasculation,  pulsatility index of arteria uterina, and endometrium height, respectively. Maximum likelihood   
leads to infinite maximum likelihood estimate of $\beta_2$ due to  quasi-complete separation. Let us consider $\beta_2$ as the parameter of interest while the remaining regression coefficients  are treated as nuisance parameters. Both  $\hat\beta^*_{2}$ and $\tilde\beta_{2_P}$ from (\ref{modprofscore}) are finite with $\hat\beta^*_{2}=2.929$ and $\tilde\beta_{2_P}=3.883$. The corresponding standard errors are 1.551 and 2.407, respectively. 
 
To assess the properties of   estimators of $\beta_{2}$, we performed  a simulation study with  sample size and covariates as in the endometrial dataset and with $\beta=(1.5,2,0,-2)^\T$. The results are presented in Table \ref{tab8b}. We found  684 samples out of 10,000 with data separation. Empirical probability of underestimation indicates that $\tilde\beta_{2_P}$ has a remarkable performance in terms median centering. On the other hand, as expected, $\hat\beta^*_{2}$ has estimated bias close to zero. Coverages of Wald-type confidence intervals based on $\hat\beta^*_{2}$ and on $\tilde\beta_{2_P}$ are comparable, while those based on $\hat\beta_2$ are favoured by being computed only using samples with finite estimates. Score-type intervals  based on  $\tilde{U}_{_P}(\beta_2)$ perform slightly better than Wald-type ones, while score-type confidence intervals for scalar components of the parameter are not available when using bias reduction.

\begin{table}\centering
 \begin{threeparttable}
 \def~{\hphantom{0}}
\caption{Simulation results for endometrial cancer study. 
%Simulation of  estimates of $\beta_{1}$ using maximum likelihood $\hat\beta_{1}$, bias reduction $\hat\beta_{1}^*$ and median bias reduction $\tilde\beta_{1_P}$. 
For $\hat\beta_{2}$, 
B, RMSE and coverage  are conditional upon finiteness of the estimates}{
\begin{tabular}{lcccccc}
 &                        PU  &    MAE  &   B      &     RMSE         & Wald   & Score \\   
 $\hat\beta_{2}$ &      43.0  &    0.66 &   0.12    &     0.90         &   97.5 & 98.9 \\ 
   $\hat\beta^*_{2}$ &  53.1  &    0.56 &   0.02    &     0.90         &    97.4 & --\\%96.5 \\ 
   $\tilde\beta_{2_P}$ &  49.7  &    0.60 &    0.16   &     1.09         &    97.7 & 95.7\\ 
\end{tabular}}
\label{tab8b}
\begin{tablenotes}\footnotesize\item
PU, percentage of underestimation; MAE,   median absolute error; B, bias; RMSE, root mean squared error; Wald and Score, percentage coverage of 95\% Wald-type and score-type confidence intervals.  
\end{tablenotes}
\end{threeparttable}
\end{table}
\end{example}

 The modified profile score (\ref{modprofscore}) can also be seen as a median modification of a first order bias corrected profile score $U_{_P}(\psi)-\kappa_{1\psi}$, with $\kappa_{1\psi}$ evaluated at $(\psi, \hat\lambda_\psi)$ \citep{mccu1990}. This is equivalent to the score of an adjusted profile likelihood, such as the modified profile likelihood \citep{Barndorff-Nielsen:1983}. Many available adjustments of the profile likelihood share indeed  the common feature of reducing the score bias to $O(n^{-1})$ \citep{diciccioetal1996}. In the presence of many nuisance parameters, typically the term $\kappa_{1\psi}$ dominates $\kappa_{3\psi}/\kappa_{2\psi}$. For instance, in a stratified setting with independent $y_{aj}$, $a=1,\ldots,q$ and $j=1,\ldots,m$, having marginal distribution depending on $(\psi,\lambda_a)$ with both $q$ and $m$ diverging as in   \citet{sart2003},  the term $\kappa_{1\psi}$ 
in (\ref{modprofscore}) is of order $O(q)$, while $\kappa_{3\psi}/\kappa_{2\psi}$ is of order $O(1)$. Therefore, the difference between $\tilde\psi_{_P}$ and $\hat\psi_{_M}$, the maximizer of the modified profile likelihood, is of order $O\{1/(qm)\}$ and both estimators have  the standard asymptotic behaviour provided that $q=o(m^3)$, as opposed to the stronger condition $q=o(m)$ for the maximum likelihood estimator. 

\begin{example}
 Gamma samples with common shape parameter.
 Let $y_{aj}$, $a=1,\ldots,q$ and $j=1,\ldots,m$, be realizations of independent gamma random variables with  shape parameter $\psi$ and scale parameter $1/\lambda_a$. 
 The needed quantities in (\ref{modprofscore}) are 
\begin{equation*}
\begin{split}
U_{_P}(\psi) &= t+ qm\log m\psi-m\Psi^{(0)}(\psi), \quad \nu_{\psi,\psi}=mq\Psi^{(1)}(\psi),\quad \nu_{\psi,a}=-m/\lambda_a,\\
 \nu_{a,a}&=(m\psi)/\lambda^2_a, \quad \nu_{a,\psi,\psi}=0,  \qquad  \nu_{a,b}=\nu_{a,b,\psi}=
 \nu_{a,b,c}=0, \quad  a\neq b, \\
  \nu_{\psi,\psi,\psi}&= mq\Psi^{(2)}(\psi),\quad \nu_{a,a,\psi}=m/\lambda^2_a,\quad \nu_{a,a,a}=-(2m\psi)/\lambda^3_a,
\end{split}
\end{equation*}
where $t=\sum_{a=1}^q\sum_{j=1}^m\log y_{aj}$ and $\Psi^{(k)}(\psi)=d^{k+1}\log\Psi(\psi)/d\psi^{k+1} $ is the poly-gamma function of order $k.$ 
Here, the conditional maximum likelihood estimator, $\hat\psi_{_C}$, based on the   distribution of $t$ given the stratum sums is also available and is  asymptotically equivalent to both  $\tilde\psi_{_P}$ and    $\hat\psi_{_M}$, provided that $q=o(m^3)$ \citep[][Example 2]{sart2003}.

Simulation results with 10,000 replications   are shown in Table \ref{tab345} for $q=1,50$, $m=5,10$, $\psi=\exp(1)$. We compared  $\hat\psi$,   $\hat\psi_{_M}$, $\hat\psi_{_C}$,   $\tilde\psi_{_P}$, the bias reduced estimator $\hat\psi^*$ in the $(\psi,\lambda)$ parameterization and the estimator $\hat\psi^{**}=\exp(\hat\varphi^*)$, where  $\hat\varphi^*$ is  the bias reduced estimator of $\varphi$ in the  parameterization $(\varphi,\chi)$, with $\varphi=\log\psi$, $\chi=\log\lambda$.  Median centering of $\tilde\psi_{_P}$ is considerable, even in the most extreme setting with $q=50$. 
Median bias reduction shows coverage of Wald-type confidence intervals closer to nominal values than bias reduction. Score-type intervals based on $\tilde{U}_{_P}(\psi)$ are slightly more accurate than Wald-type ones. As expected, the $\varphi$ parameterization is more favourable than the $\psi$ parameterization for   bias reduction.  
%As in Example \ref{ex-SN},  the effect of median bias reduction is milder than that of bias reduction.  

\begin{table}\centering
 \begin{threeparttable}
 \def~{\hphantom{0}}
\caption{Simulation results for estimates of the common gamma shape parameter
%, using maximum likelihood $\hat\psi$, modified profile maximum likelihood $\hat\psi_{_M}$,
%conditional maximum likelihood $\hat\psi{_C}$, bias reduction $\hat\psi^*$, reparameterized bias reduction $\hat\psi^{**}$ 
%and median bias reduction $\tilde\psi_{_P}$
}
\begin{tabular}{cclcccccc}
   % \toprule
$q$&$m$&&                    PU &MAE  &    B &  RMSE  & Wald   & Score   \\   
1& 5&$\hat\psi$  & 		     29.9 &1.41 & 3.48 & 8.36&  97.5& 97.5\\
& & $\hat\psi_{_M}$&          40.9& 1.22 &  2.31  &6.51 &    95.2  &    95.1\\
 & & $\hat\psi{_C}$ &  		41.0 &1.22 & 2.30 & 6.51&  95.1& 95.0\\
 & &$\hat\psi^*$ &  		73.4 &1.27 & -0.04 &2.99 &  76.1&--\\
 & &$\hat\psi^{**}$ & 		56.5 &1.22 & 1.06 & 4.68 &  84.3&--\\
 & &$\tilde\psi_{_P}$ & 		     50.1 &1.19 & 1.51 & 5.29 &  89.2 &94.9\\
 & 10&$\hat\psi$ &  		35.7 &0.85 & 1.03 & 2.36 &  97.1& 97.1\\
& & $\hat\psi_{_M}$&  44.3  & 0.80 & 0.68 & 2.03  &   95.6 &     95.5\\
& &$\hat\psi{_C}$& 		     44.3 &0.80 & 0.68 & 2.03 &  95.5& 95.5\\
&&$\hat\psi^*$ & 		     64.5 &0.85 & -0.03 &1.48 &  85.6&--\\
 &&$\hat\psi^{**}$ & 		54.7 &0.80 & 0.31 & 1.73 &  91.2&--\\
& &$\tilde\psi_{_P}$  & 		     50.5 &0.79 & 0.45 & 1.83 &  92.8& 95.3\\
50&   5&$\hat\psi$& 		1.2  &0.62 & 0.64 & 0.72 &  40.5& 40.5\\
& & $\hat\psi_{_M}$&        47.0 & 0.17 &  0.03 & 0.26 &    95.1 &     95.1\\
& & $\hat\psi{_C}$ & 		     48.3 &0.17 & 0.03 & 0.26 &  95.0&95.0\\
 & &$\hat\psi^*$ & 		     58.0 &0.18 & -0.04 &0.26 &  90.2&--\\
 & &$\hat\psi^{**}$ & 		51.2 &0.06 & 0.00 & 0.09 &  92.2&--\\
 & &$\tilde\psi_{_P}$ &  		48.4 &0.17 & 0.03 & 0.26 &  92.5&97.1\\
& 10&$\hat\psi$& 		     6.2  &0.27 & 0.28 & 0.34 &  67.8&67.8\\
& & $\hat\psi_{_M}$&       48.6 & 0.11 & 0.01 &0.17 &    95.1 &     95.1\\
&&$\hat\psi{_C}$ &  		     49.0 &0.11 & 0.01 & 0.17 &  95.1&95.1\\
 & &$\hat\psi^*$ & 		     53.6 &0.12 & -0.01 &0.17 &  93.4&--\\
&  &$\hat\psi^{**}$ &		50.5 &0.04 & 0.00 & 0.06 &  93.9&--\\
 & &$\tilde\psi_{_P}$ & 		     49.5 &0.12 & 0.01  &0.17 &  94.0&96.0\\
 \end{tabular}
\label{tab345}
\begin{tablenotes}\footnotesize\item
PU, percentage of underestimation; MAE,   median absolute error; B, bias; RMSE, root mean squared error; Wald and Score, percentage coverage of 95\% Wald-type and score-type confidence intervals.  
\end{tablenotes}
\end{threeparttable}
\end{table}
 \end{example}

\begin{example} Common odds ratio in $2\times 2$ tables.
% spostare nel paragrafo successivo (si userebbe la componente 'odds ratio' della stima median unbiased globale
Consider $q$ independent pairs of observations $(y_{a1},y_{a2})$, realizations of independent binomial variables $Bi(1,p_{a1})$
and $Bi(m,p_{a2})$. Let $p_{a1}=\exp(\lambda_a+\psi)/\{1+\exp(\lambda_a+\psi)\}$ and 
$p_{a2}=\exp(\lambda_a)/\{1+\exp(\lambda_a)\}$. 
This model may arise in case-control
studies, with  1 case and $m$ controls in each table, and
where interest is about $\psi$, representing the influence of some risk factor.  
As in \citet{breslow81}, we consider sparse settings with large $q$ and small $m$, where improvements over the  maximum likelihood estimator are particularly needed. This is also an instance where invariance is important, since results are often reported in terms of odds ratio $\rho=\exp(\psi)$. The median modified profile score is a special case of that in Example \ref{binaryregression}.    The conditional maximum likelihood estimator $\hat\psi_{_C}$ is available, based on the conditional distribution of $t=\sum_{a=1}^q y_{a1}$ given  $s_a=y_{a1}+y_{a2}$, $a=1,\ldots,q$.

The aim is to compare the various methods  with conditional maximum likelihood, which gives  consistency also for fixed $m$ and can be considered as a gold standard. The comparison is made on the odds ratio scale, using $\hat\rho^*=\exp(\hat\psi^*)$ for bias reduction and equivariance for the other estimators. As in \citet[][Example 3]{sart2003},  we focus on  particular instances with odd values of $m$ and with $s_a=(m+1)/2$, $a=1,\ldots,q$, so that all tables have the same  number of successes and failures.   In this case, $\hat\rho_{_C}=(t/q)/(1-t/q)$ and, for a given $m$, both $\hat\rho$ and $\hat\rho_{_M}$ are functions of  $t/q$ only. Although $\hat\rho^*$ and $\tilde\rho_{_P}$ depend also on $q$,  numerical evidence indicates that such dependence vanishes as $q$ increases. For $q=300$ and various values of $m$, estimates of the odds ratio are plotted versus $\hat\rho_{_C}$ in Figure \ref{fig-odds}. Median modified estimates are almost indistinguishable from those based on modified profile likelihood, as expected, and both are the closest to  $\hat\rho_{_C}$. On the contrary, $\hat\rho$ markedly departs from $\hat\rho_{_C}$, especially for small $m$ and as  $\hat\rho_{_C}$ increases, while $\hat\rho^*$ overcorrects, in particular for large values of  $\hat\rho_{_C}$. Other values of $q$ give the same results in terms of estimates, while accuracy of inference is affected since standard errors decrease as $q$ increases.  

\begin{figure}\centering
%\vspace{-1cm}
\includegraphics[width=0.95\textwidth]{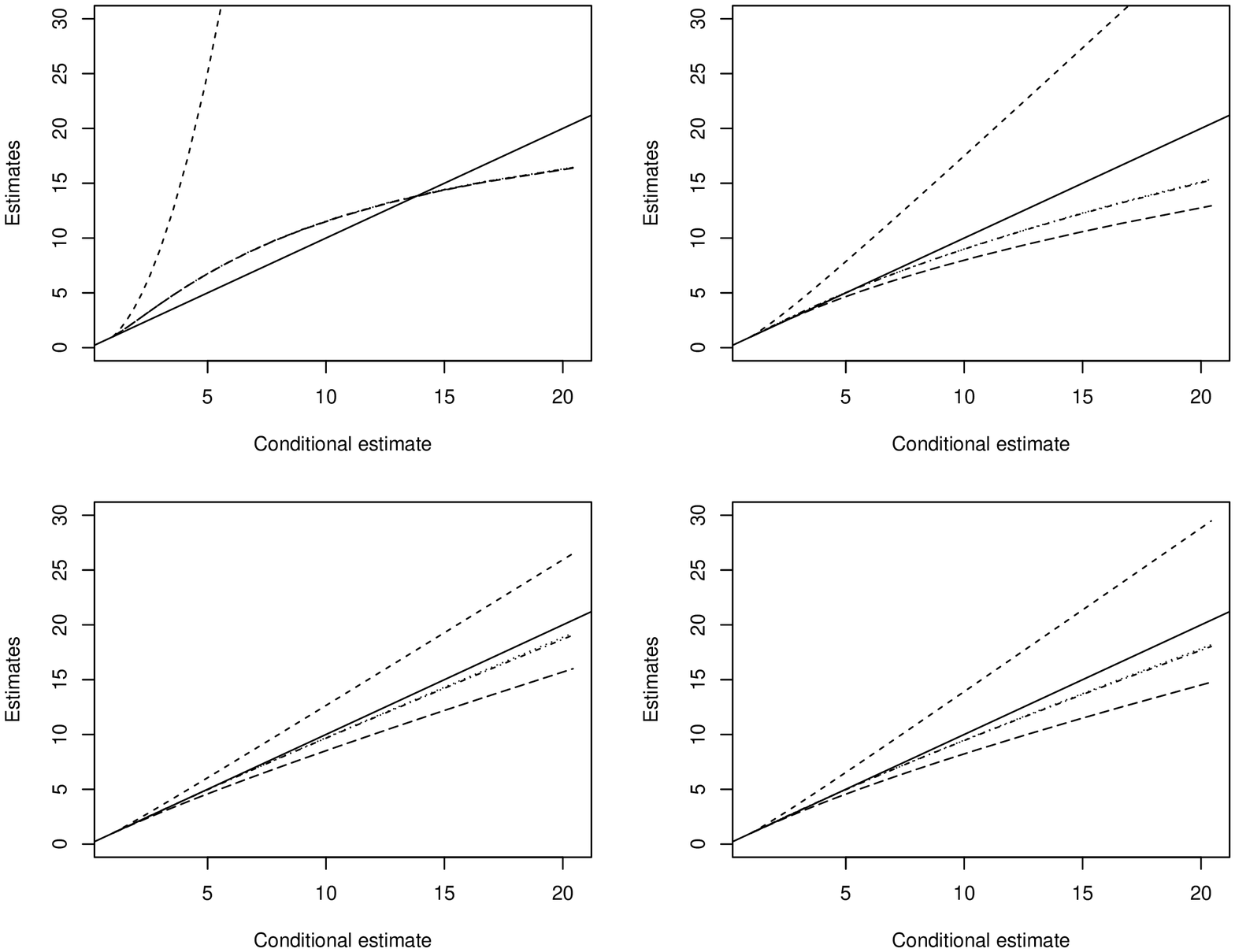}
%\vspace{-1.cm}
\caption{Estimates of odds ratio as functions of the conditional maximum likelihood estimate for $q=300$ and $m=1,3,5,7$ (clockwise from top left): $\hat\rho_{_C}$ (solid), $\hat\rho$ (dashed), $\hat\rho^*$ (long-dashed),  $\hat\rho_{_M}$ (dotted), $\tilde\rho_{_P}$ (dot-dashed).}
\label{fig-odds}
\end{figure}

\end{example}

%%%%%%%%%%%%%%%%%%%%%%%%%%%%%%%%%%%%%%%%%%%%%%%%%%%%%%%%%%%%%%%%
\section{Median modified score for a vector parameter}\label{vector}
%%%%%%%%%%%%%%%%%%%%%%%%%%%%%%%%%%%%%%%%%%%%%%%%%%%%%%%%%%%%%%%%
For estimation of the full vector parameter $\theta$, with $p>1$, a direct extension 
of the rationale leading to  (\ref{modscore}) does not seem to be practicable due to lack of a manageable definition of multivariate median. Actually, a number of definitions have been proposed  \citep{oja2013}, but  none seems suitable for developing  a median modification of the score vector. For instance, with the simplest definition, i.e.\ taking the vector of approximate marginal medians as an approximate median of the score vector, dependence among score components is ignored. 
%Indeed, unreported  simulation results show that the resulting corrected estimator has rather poor sampling properties.  
Other available definitions of multivariate median would involve the joint distribution of the score vector in a rather complex way and do not seem to  provide feasible proposals. 

Instead, the approach we follow is to set up a system of estimating equations giving, for each $\theta_r$, $r=1,\ldots,p$, the same estimate as (\ref{modprofscore}), up to terms of order $O_p(n^{-1})$ included. This is obtained by defining the median modified score vector $\tilde{U}(\theta)$ with components   
\begin{equation}\label{jointmed}
\tilde{U}_r=U_r-\gamma_{ra}U_a +M_r \,, \qquad
%-\kappa_{1r}+\frac{1}{6} \frac{\kappa_{3r}}{\kappa_{2r} }\,, \qquad 
r=1,\ldots,p\,,
 \end{equation}
where $M_r= -\kappa_{1r}+\kappa_{3r}/(6\kappa_{2r})$, and $\kappa_{jr}$, $j=1,2,3$, are as in (\ref{kprof}) with $\psi=\theta_r$. 
In (\ref{jointmed}), and in related formulae (\ref{kprof}), indices $a,b\ldots$ take values in $\{1,\ldots,p\}\setminus \{r\}$, and are summed when repeated. Moreover, all quantities involved are evaluated at $\theta$, so that no constrained estimates are involved. Then, the joint estimate $\tilde\theta$ is defined as solution of $\tilde{U}(\theta)=0$.

For each $r=1,\ldots,p$, $\tilde U_r$ behaves tensorially under interest respecting reparameterizations of $\theta_r$. As a consequence, $\tilde\theta$ 
is equivariant under joint reparameterizations that transform each component of $\theta$ separately.

%Denoting by  $\bar{U}(\theta)$ the vector with components given by the efficient scores  $\bar{U}_r= U_r-\gamma_{ra} U_a $,  equation
%$\bar{U}(\theta)=0$ has the same solution as $U(\theta)=0$,  namely the maximum likelihood estimate $\hat\theta$.  Indeed, we can write $\bar{U}(\theta)=A(\theta) U(\theta)$, with $A(\theta)$ a nonsingular and nonstochastic matrix  of order $p$ which can be expressed as a function of $i(\theta)^{-1}$.  As shown in the Appendix,  $H(\theta)=E_{\theta}(-\partial \bar{U}(\theta)/\partial\theta^\T)=\{\diag(i(\theta)^{-1})\}^{-1}$. Moreover,   $H(\theta)=A(\theta)i(\theta)$, so that $A(\theta)=  H(\theta) i(\theta)^{-1}$. 
%Since $\tilde{U}(\theta)=\bar{U}(\theta)+O(1)$, it follows that $\tilde\theta$ differs from $\hat\theta$ by $O_p(n^{-1})$ and  the asymptotic distribution of $\tilde\theta$ is the same as that of the maximum likelihood estimator $\hat\theta$. 

Denoting by  $\bar{U}(\theta)$ the vector with components given by the efficient scores  $\bar{U}_r= U_r-\gamma_{ra} U_a $,  we can write $\bar{U}(\theta)=A(\theta) U(\theta)$, with $A(\theta)$ a nonsingular and nonstochastic matrix  of order $p$. % which can be expressed as a function of $i(\theta)^{-1}$.
 As shown in (\ref{Hessian}),   $H(\theta)=E_{\theta}\left\{-\partial \bar{U}(\theta)/\partial\theta^\T\right\}=\{\diag(i(\theta)^{-1})\}^{-1}$. Moreover,   $H(\theta)=A(\theta)i(\theta)$, so that $A(\theta)=  H(\theta) i(\theta)^{-1}$. 
Hence, solving $\tilde U(\theta)= 0$  is equivalent to solving
\begin{equation} \label{jointmue2}
U(\theta) + i(\theta) M_1(\theta) =0\,,
\end{equation}
with $M_1(\theta)$ having elements $M_{1r}=M_r/\kappa_{2r}$. There is no general guarantee that (\ref{jointmue2}) has a solution. However,  $i(\theta) M_1(\theta)$ is of order $O(1)$, so that, asymptotically,  existence of $\tilde\theta$ is guaranteed whenever $\hat\theta$ exists. Moreover,  
$\tilde\theta -\hat\theta=O_p(n^{-1})$ and  the asymptotic distribution of $\tilde\theta$ is the same as that of  $\hat\theta$.

Let $\tilde\theta_r$ be the $r$-th component of $\tilde\theta$  and   $\tilde\theta_{r_P}$  the solution of $\tilde{U}_{_P}(\theta_r)=0$, with $\tilde{U}_{_P}(\cdot)$ given by (\ref{modprofscore}).  
In a regular model, we show that
\begin{equation}\label{equiv}
 \tilde\theta_r-\tilde\theta_{r_P}=O_p(n^{-3/2})\,,
\end{equation}
$r=1,\ldots,p$. A  proof of (\ref{equiv}) is given in the Appendix. 
A key property for the result is that $H(\theta)$ is a diagonal matrix, so that $\tilde{U}(\theta)$ satisfies
%under interest respecting reparameterizations of $\theta_r$, that is 
\begin{equation}\label{S.orth}
E_{\theta}(\partial \tilde U_r/\partial\theta_s )=O(1), \quad r,s=1,\ldots, p, \quad s\neq r\,.
\end{equation}
Following \citet{jorgensenknudsen2004},   we call $\tilde U_r$  first-order insensitive to $\theta$ components other than $\theta_r$, $r=1,\ldots,p$. Due to (\ref{S.orth}), terms up to order $O_p(n^{-1})$ in the expansion of $\tilde\theta_r-\theta_r$ are not affected by terms of order $O(1)$ in $\tilde U_s$,  $s\neq r$.   
 
Using delta method arguments as in \citet[][\S~2.7]{hall1992}, it follows from (\ref{equiv}) that, in the continuous case,  
$\P_{\theta}( \tilde\theta_r \leq  \theta_r)= \P_{\theta}( \tilde\theta_{r_P} \leq  \theta_r) +O(n^{-3/2})$, so that componentwise median unbiasedness of $\tilde\theta$ with error of order $O(n^{-3/2})$ follows from the analogous property of $\tilde\theta_{r_P}$.  

Equation (\ref{jointmue2}) has the same structure as the estimating equation for bias reduction. Hence, some of the ideas in \citet{kosmi2009,kosmi2010} for the implementation of  bias reduction  can be adapted for median bias reduction. For instance, a modified Fisher scoring iteration can be written as 
\begin{equation}\label{fisherscoring}
\tilde\theta^{(k+1)} = \tilde\theta^{(k)} + M_1(\tilde\theta^{(k)}) + i(\tilde\theta^{(k)})^{-1} U(\tilde\theta^{(k)})\,,
\end{equation}
which differs from the analogue for $\hat\theta$ only by the addition of the term $M_1$. When available, $\hat\theta$ is a convenient starting value.  %(see also (\ref{expprofilemue})). 
As happens for bias reduction \citep{kosmi2010}, convergence or otherwise of  (\ref{fisherscoring}) depends on the properties of the specific assumed model. Nonetheless, assuming convergence of (\ref{fisherscoring}), it will be to a solution of (\ref{jointmue2}).

\begin{example} Binary regression (continued). Quantities needed for (\ref{jointmed}) in binary regression are the same as those in Example \ref{binaryregression}. Moreover,   (\ref{fisherscoring}) simplifies to a modified iterative reweighted least squares procedure.
Details are provided in the Supplementary Material and an implementation is given in the {\tt R} package {\tt mbrglm} \citep{kpal2017a}. 

For the endometrial cancer grade dataset, estimates of the model parameters using  (\ref{jointmed}) for logistic and probit regression are given in Table \ref{kenne:tab2}. The estimate  $\tilde\beta_2$ is very close to $\tilde\beta_{2_P}$ obtained  in Example \ref{binaryregression} as a solution of  (\ref{modprofscore}).

\begin{table}\centering
 \def~{\hphantom{0}}
\caption{Endometrial cancer study. Estimates (s.e.) for logistic regression (top rows) and probit regression (bottom rows)}{
\begin{tabular}{lcccc}
& $\beta_1$ & $\beta_2$& $\beta_3$ & $\beta_4$ \\ 
% \midrule
   $\hat\beta$ & 4.305 (1.637) & $+\infty$ $(+\infty)$ & -0.042 (0.044) & -2.903 (0.846) \\ 
   $\hat\beta^*$  & 3.775 (1.489) & 2.929 (1.551) & -0.035 (0.040) & -2.604 (0.776) \\ 
    $\tilde\beta$  & 3.969 (1.552) & 3.869 (2.298) & -0.039 (0.042) & -2.708 (0.803) \\ \\
       $\hat\beta$  & 2.181  (0.857) & $+\infty$ $(+\infty)$ & -0.019  (0.024)& -1.526 (0.433)\\
   $\hat\beta^*$ & 1.915  (0.789) & 1.659 (0.747)& -0.015 (0.021) & -1.380 (0.403)\\ 
   $\tilde\beta$ & 1.984 (0.812) & 1.971 (0.919) & -0.017 (0.022)& -1.425 (0.414) \\ 
\end{tabular}}
\label{kenne:tab2}
\end{table}

The same simulated samples as in Example \ref{binaryregression} allow to evaluate the properties of estimators of the vector $\beta$. 
Table \ref{kenne:tab1} shows  that the new method is remarkably accurate in achieving median centering for all the parameter components. It should be recalled that  684 samples out of 10,000 produced infinite maximum likelihood estimates, so that results for $\hat\beta$ should be judged accordingly. 
The three approaches  are comparable in terms of coverage of  Wald-type confidence intervals, while profile score-type intervals show some improvement. 
%As expected the bias corrected
%estimator of \citet{firth1993} has good performance in terms of bias.
Similar results have been found with a probit model and are reported in the Supplementary Material.
% Table \ref{endoall2} displays parameter estimates while simulation results are summarized in Table \ref{endosimall2}.

\begin{table}\centering
 \begin{threeparttable}
 \def~{\hphantom{0}}
\caption{Simulation results for endometrial cancer study. 
%Simulation of estimates of the regression coefficients with logistic link.  
For  maximum likelihood,  B, RMSE and coverage are conditional upon finiteness of the estimates}{
\begin{tabular}{lcccccc}
              &             PU &       MAE &         B &     RMSE & Wald  & Score \\                                      
$\hat\beta$  &             45.1 &       0.97&      0.29   &    1.60 &       95.8 & 94.8\\                         
             &             43.0 &       0.66&      0.12   &    0.90 &       97.4  &95.2\\                         
             &             51.0 &       0.03 &     0.00    &   0.04  &       95.0  &94.2\\                         
             &             56.0 &       0.57 &     -0.26    &  1.02  &       96.0  &94.9\\                          
$\hat\beta^*$&             52.6 &       0.86&      0.00   &    1.38 &       96.6 & --\\%95.9\\                         
             &             53.0 &       0.56&      0.02   &    0.90 &       97.4  & --\\%96.5\\                         
             &             49.6 &       0.02 &     0.00    &   0.04  &       96.3 &--\\%95.1\\                         
             &             44.4 &       0.52&      0.01   &    0.83 &       94.8 &--\\%96.2\\                            
$\tilde\beta$&             50.1 &       0.90&      0.09   &    1.46 &       96.4 & 95.0\\                         
             &             49.7 &       0.59&      0.15   &    1.07 &       97.5  &95.3\\                         
             &             50.7 &       0.02 &     0.00    &   0.04  &       96.1 &94.3\\                         
             &             49.6 &       0.52 &     -0.10    &  0.89  &       95.8  &94.7\\ 
\end{tabular}}
\label{kenne:tab1}
%\begin{tablenotes}
%      \small
%      \item For the mle, 85 samples  without convergence
%      \item For the  mean bias correction, 65 samples
%      \item  For the median bias correction 82 samples
%    \end{tablenotes}
\begin{tablenotes}\footnotesize\item
PU, percentage of underestimation; MAE,   median absolute error; B, bias; RMSE, root mean squared error; Wald and Score, percentage coverage of 95\% Wald-type and score-type confidence intervals.  
\end{tablenotes}
\end{threeparttable}
\end{table}
\end{example}

\begin{example} Beta regression.   
Let $y_i$, $i=1,\ldots,n$, be independent realizations of beta random variables with parameters $\phi\mu_i$ and $\phi(1-\mu_i)$, i.e. with expected value $\mu_i$ and precision parameter $\phi$. We assume a  regression structure for the expected value $\mu_i=g^{-1}(\eta_i)$, 
%and precision parameter $\phi$, 
where $\eta_i= x_i \beta$, $x_i=(x_{i1},\ldots, x_{ip})$ is a vector of covariates and $g(\cdot)$ is a given link function, such as the logit.  The needed quantities for  (\ref{jointmed}) with $\theta=(\beta_1,\ldots,\beta_p,\phi)^\T$ are the same as those required for bias reduction \citep{kosmi2010}. Details are given in the Supplementary Material. An 
\texttt{R} implementation of (\ref{fisherscoring}) is given in function 
\texttt{mbrbetareg}, available on {\tt GitHub}. %\citep{kpal2017b}.

As an application, we consider data in \citet[Table 15.4]{Grif1993} on food expenditure for a random sample of 38 households in a large U.S. city,  also available in the {\tt R}  package {\tt betareg}.  The objective is to model the proportion
of income spent on food ($y$) as a function of  income ($x_2$) and number of persons ($x_3$).  Estimates of $\theta=(\beta_1,\beta_2,\beta_3,\phi)^\T$, where $\beta_1$ is an intercept, and with the logit link are given in Table \ref{kenne:tabfood1}. Values for the regression coeffcients and corresponding standard errors are essentially the same for all methods, while differences are observed for the dispersion parameter.

 \begin{table}
\def~{\hphantom{0}}
\caption{Food expenditure. Estimates (s.e.) for beta regression}{
\begin{tabular}{lcccc}
   & $\beta_1$ &$\beta_2$  & $\beta_3$ & $\phi$ \\ 
$\hat\theta$ & -0.623  (0.224) & -0.012 (0.003) & 0.118 (0.035) & 35.610 (8.080) \\ 
 $ \hat\theta^*$ & -0.621  (0.239) & -0.012 (0.003) & 0.118  (0.038) & 30.922 (7.005) \\ 
  $\tilde\theta$  & -0.621 (0.235) & -0.012 (0.003) & 0.118 (0.037) & 32.160 (7.289) \\ 
\end{tabular}}
\label{kenne:tabfood1}
\end{table}

We performed a simulation study with the same sample size and covariates as in the food expenditure data and with parameter fixed at $\hat\theta$.  Results obtained from 100,000 simulated samples show identically accurate behaviour for estimators of regression parameters. Hence, only results for estimators of $\phi$ are displayed in Table \ref{kenne:tabfood2}, in line with those of previous examples. The complete table is reported in the Supplementary Material, together with an additional example with a smaller ratio $n/p$ leading to larger differences among estimators of $\phi$. This also implies different confidence intervals for the   regression coefficients and corresponding coverages. 

\begin{table}\centering
 \begin{threeparttable}
 \def~{\hphantom{0}}
\caption{Simulation results for food expenditure}{
\begin{tabular}{lcccccc}
 & PU & MAE & B & RMSE & Wald  & Score \\ 
%   &  50.0 & 0.15 & 0.00 & 0.22 & 93.1 & 94.3\\
%  & 50.6  & 0.00 & 0.00 & 0.00 & 93.5 & 94.8\\
% &  50.5 &  0.02 & 0.00  & 0.04 & 93.2 & 94.5\\
 $\hat\phi$&   32.7 & 6.25 & 5.46 & 11.79 & 95.1 & 95.1\\
% & 49.8 & 0.15 & 0.00 & 0.22 & 94.7 & --\\%96.2\\
%& 50.2 & 0.00 & 0.00 & 0.00 & 95.2 & --\\%96.3\\
%& 50.9 & 0.02 & 0.00 & 0.04 & 94.9 & --\\% 96.2\\
$\hat\phi^*$& 56.5 & 5.74  & 0.06  & 9.07 & 91.8 & --\\%95.1\\
%& 49.9 & 0.15 & 0.00 & 0.22 & 94.3 & 94.4\\
%& 50.3 & 0.00 & 0.00 & 0.00 & 94.8 & 94.8\\
%&  50.8 & 0.02 & 0.00 & 0.04 & 94.5 & 94.5\\
$\tilde\phi$&  49.8 & 5.69 & 1.49 & 9.56 & 93.7 & 95.8\\
\end{tabular}}
\label{kenne:tabfood2}
\begin{tablenotes}\footnotesize\item
    PU, percentage of underestimation; MAE, median absolute error; B, bias; RMSE, root mean squared error; Wald and Score, percentage coverage of 95\% Wald-type and score-type confidence intervals.   
\end{tablenotes}
\end{threeparttable}
\end{table}

\end{example}

\appendix
%\appendixone
\section*{Appendix}
%\label{App1}
%\subsection*{Proof of (\ref{mu_modscore})}
{\it Proof of (\ref{mu_modscore})}. Let $\rho_3=\nu_{\theta,\theta,\theta}/i(\theta)^{3/2}$ be the third standardized cumulant of $U(\theta)$, of order $O(n^{-1/2})$.
%, and let $\Phi(\cdot)$ and $\phi(\cdot)$ denote the standard normal distribution function and density. 
Then, with a standard Edgeworth expansion,  
\begin{eqnarray*}
\P_{\theta}\{\tilde{U}(\theta)\leq 0 \}&=& \P_{\theta}\{U(\theta) + \rho_3\sqrt{i(\theta)/6}\leq 0 \}
= \P_{\theta}\{U(\theta)/\sqrt{i(\theta)}\leq  -\rho_3/6 \}\\
&=& \Phi(-\rho_3/6) - \phi(-\rho_3/6)\{ \rho_3/6(\rho_3^2/36-1) +O(n^{-3/2})\} \,,
\end{eqnarray*}
where the error is of order $O(n^{-3/2})$ because the $O(n^{-1})$ term in the Edgeworth expansion is a linear combination with coefficients of order $O(n^{-1})$ of odd Hermite polynomials evaluated at $-\rho_3/6$. The result in (\ref{mu_modscore}) follows using the expansions $\Phi(-\rho_3/6)=1/2-\rho_3\phi(0)/6+O(n^{-3/2})$ and $\phi(-\rho_3/6)= \phi(0)+O(n^{-1})$. \\

%\appendixtwo
%\section*{Appendix 2}
%\label{App2}
%\subsection*{Proof of (\ref{equiv})}
{\it Proof of (\ref{equiv})}. First, an expansion of $\tilde\theta_{r_P}-\theta_r$ is readily obtained from an expansion for $\hat\theta_r-\theta_r$ \citep[see e.g.][formula (9.61)]{pace:salv:1997}, taking into account the effect of the modification to the profile score $U_{_P}(\theta_r)$ given in (\ref{modprofscore}). In detail, being 
$
-\partial \tilde{U}_{_P}(\theta_r)/\partial\theta_r=\kappa_{2r}+O_p(n^{1/2})\,,
$
we get
\begin{equation}\label{expprofilemue}
\tilde\theta_{r_P}-\theta_r=\hat\theta_r-\theta_r 
-\kappa_{1r}/\kappa_{2r}+ \kappa_{3r}/\{6\,\kappa_{2r}^2\}+O_p(n^{-3/2})\,.
\end{equation}

Second, an expansion for $\tilde\theta_r-\theta_r$ from (\ref{jointmed})  is obtained using standard asymptotic expansions for estimating equations. Let $g=g(\theta)=g(\theta;y)$ be an estimating function with generic component $g_r$.  We assume that $g$ is of order  $O_p(n^{1/2})$ with expected value $O(1)$. Let $g_{r/s}=\partial g_r/\partial\theta_s$,  
$g_{r/st}=\partial^2 g_r/(\partial\theta_s\partial\theta_t)$ and let $\xi_r=E_{\theta}(g_r)=O(1)$, $\xi_{r/s}=E_{\theta}(g_{r/s})$, $\xi_{r/st}=E_{\theta}(g_{r/st})$, the latter two quantities being typically of order $O(n)$. Moreover, let $D_{r/s}=g_{r/s}-\xi_{r/s}$, $D_{r/st}=g_{r/st}-\xi_{r/st}$. Let $\tau^{rs}$ be a generic entry of the inverse of the matrix with entries $-\xi_{r/s}$. An asymptotic expansion for $g(\bar\theta)=0$ gives
\begin{equation}\label{0.b}
\bar\theta_r-\theta_r=\tau^{rs}g_s+\tau^{rs}\tau^{tu}D_{s/t} g_u
+\frac{1}{2} \tau^{rs}\tau^{tv}\tau^{uw}\xi_{s/tu}g_v g_w +O_p(n^{-3/2})\,.
\end{equation}

When $g_r= U_r$, we obtain $\bar\theta=\hat\theta$, so that expansion (\ref{0.b}) gives the usual expansion for $\hat\theta_r-\theta_r$. The same is true if $g_r=\bar U_r$, being  $\bar U_r$ a linear transformation of $ U_r$. However, in the latter case, $\tau^{rs}=0$ if $r\neq s$, while $\tau^{rr}=i^{rr}=\kappa_{2r}^{-1}$. Indeed,
\begin{equation}\label{Hessian}
\xi_{r/s}=E_{\theta}(\bar U_{r/s})=E_{\theta}( U_{rs}-\gamma_{ra} U_{as}-
\gamma_{ra/s} U_{a})=-(i_{rs}-\gamma_{ra}i_{as})\,.
\end{equation}
Since, when $s\neq r$, we have $\nu^{ab}i_{as}= \delta^b_s$, the indicator of $b=s$, 
it follows that  $\xi_{r/s}=-(i_{rs}-i_{rb}\nu^{ab}i_{as})=-(i_{rs}-i_{rb}\delta^b_s)=0$ if $s\neq r$. On the other hand, $\xi_{r/r}=-\kappa_{2r}$. 

When (\ref{0.b}) is applied to (\ref{jointmed}), we have
$\tau^{rr}=i^{rr}+O(n^{-2})$ and $\tau^{rs}=O(n^{-2})$ if $r\neq s$. Therefore, terms up to order $O_p(n^{-1})$ in the  expansion for $\tilde\theta_r-\theta_r$ do not involve modification terms of order $O(1)$ of $\tilde U_s$ with $s\neq r$. The desired expansion for $\tilde\theta_r-\theta_r$ is thus equivalently obtained from the system
$$
\tilde U_r=0,\quad \bar U_s=0, s\neq r\,.
$$
This is the same as the expansion from $\bar U_r=0, r=1,\ldots, p$, plus a $O(n^{-1})$ term given by the modification term in (\ref{jointmed}) divided by $\kappa_{2r}$. Therefore, the resulting expansion coincides with (\ref{expprofilemue}).

%E' NECESSARIO DIRE CHE LE RIMANENTI QUANTITA' ($D_{s/t}$ E $\xi_{s/tu}$) RIMANGONO INVARIATE? (la prima e' identica, la secondo differisce per un $O(1)$)

\bibliographystyle{biometrika}
\bibliography{ref.bib}

\begin{thebibliography}{29}
\expandafter\ifx\csname natexlab\endcsname\relax\def\natexlab#1{#1}\fi

\bibitem[{Agresti(2015)}]{agre2015}
\textsc{Agresti, A.} (2015).
\newblock \textit{Foundations of Linear and Generalized Linear Models}.
\newblock John Wiley \& Sons.

\bibitem[{Barndorff-Nielsen(1983)}]{Barndorff-Nielsen:1983}
\textsc{Barndorff-Nielsen, O.~E.} (1983).
\newblock {On} a formula for the distribution of the maximum likelihood
  estimator \textbf{70}, 343--365.

\bibitem[{Barndorff-Nielsen(1986)}]{barn1986}
\textsc{Barndorff-Nielsen, O.~E.} (1986).
\newblock Inference on full or partial parameters based on the standardized
  signed log likelihood ratio.
\newblock \textit{Biometrika} \textbf{73}, 307--322.

\bibitem[{Barndorff-Nielsen \& Cox(1989)}]{bncox89}
\textsc{Barndorff-Nielsen, O.~E.} \& \textsc{Cox, D.~R.} (1989).
\newblock \textit{Asymptotic Techniques for Use in Statistics}.
\newblock Chapman \& Hall.

\bibitem[{Biehler et~al.(2015)Biehler, Holling \& Doebler}]{bieh2015}
\textsc{Biehler, M.}, \textsc{Holling, H.} \& \textsc{Doebler, P.} (2015).
\newblock Saddlepoint approximations of the distribution of the person
  parameter in the two parameter logistic model.
\newblock \textit{Psychometrika} \textbf{80}, 665--688.

\bibitem[{Breslow(1981)}]{breslow81}
\textsc{Breslow, N.} (1981).
\newblock Odds ratio estimators when the data are sparse.
\newblock \textit{Biometrika} \textbf{68}, 73--84.

\bibitem[{Cai \& Wang(2009)}]{cai2009}
\textsc{Cai, T.~T.} \& \textsc{Wang, H.} (2009).
\newblock Tolerance intervals for discrete distributions in exponential
  families.
\newblock \textit{Statistica Sinica} \textbf{19}, 905--923.

\bibitem[{Cox \& Hinkley(1974)}]{cox74}
\textsc{Cox, D.~R.} \& \textsc{Hinkley, D.~V.} (1974).
\newblock \textit{Theoretical Statistics}.
\newblock Chapman and Hall, London.

\bibitem[{DiCiccio et~al.(1996)DiCiccio, Martin, Stern \&
  Young}]{diciccioetal1996}
\textsc{DiCiccio, T.}, \textsc{Martin, M.}, \textsc{Stern, S.} \&
  \textsc{Young, G.} (1996).
\newblock {Information bias and adjusted profile likelihoods}.
\newblock \textit{Journal of the Royal Statistical Society Series B}
  \textbf{58}, 189--203.

\bibitem[{Firth(1993)}]{firth1993}
\textsc{Firth, D.} (1993).
\newblock Bias reduction of maximum likelihood estimates.
\newblock \textit{Biometrika} \textbf{80}, 27--38.

\bibitem[{Giummole \& Ventura(2002)}]{gium2002}
\textsc{Giummole, F.} \& \textsc{Ventura, L.} (2002).
\newblock Practical point estimation from higher-order pivots.
\newblock \textit{Journal of Statistical Computation and Simulation}
  \textbf{72}, 419--430.

\bibitem[{Griffiths et~al.(1993)Griffiths, Hill \& Judge}]{Grif1993}
\textsc{Griffiths, W.~E.}, \textsc{Hill, R.~C.} \& \textsc{Judge, G.~G.}
  (1993).
\newblock \textit{Learning and Practicing Econometrics}.
\newblock John Wiley \& Sons.

\bibitem[{Hall(1992)}]{hall1992}
\textsc{Hall, P.} (1992).
\newblock \textit{The Bootstrap and Edgeworth Expansion}.
\newblock Springer, New York.

\bibitem[{Hirji et~al.(1989)Hirji, Tsiatis \& Mehta}]{hirji1989}
\textsc{Hirji, K.~F.}, \textsc{Tsiatis, A.~A.} \& \textsc{Mehta, C.~R.} (1989).
\newblock Median unbiased estimation for binary data.
\newblock \textit{The American Statistician} \textbf{43}, 7--11.

\bibitem[{Jorgensen \& Knudsen(2004)}]{jorgensenknudsen2004}
\textsc{Jorgensen, B.} \& \textsc{Knudsen, S.~J.} (2004).
\newblock {Parameter orthogonality and bias adjustment for estimating
  functions}.
\newblock \textit{Scandinavian Journal of Statistics} \textbf{31}, 93--114.

\bibitem[{Kenne~Pagui et~al.(2017)Kenne~Pagui, Salvan \& Sartori}]{kpal2017a}
\textsc{Kenne~Pagui, E.~C.}, \textsc{Salvan, A.} \& \textsc{Sartori, N.}
  (2017).
\newblock \textit{mbrglm: Median Bias Reduction in Binomial-Response GLMs}.
\newblock R package version 0.0.1.

\bibitem[{Kosmidis(2014)}]{kosmi2014}
\textsc{Kosmidis, I.} (2014).
\newblock Bias in parametric estimation: reduction and useful side-effects.
\newblock \textit{Wiley Interdisciplinary Reviews: Computational Statistics}
  \textbf{6}, 185--196.

\bibitem[{Kosmidis \& Firth(2009)}]{kosmi2009}
\textsc{Kosmidis, I.} \& \textsc{Firth, D.} (2009).
\newblock Bias reduction in exponential family nonlinear models.
\newblock \textit{Biometrika} \textbf{96}, 793--804.

\bibitem[{Kosmidis \& Firth(2010)}]{kosmi2010}
\textsc{Kosmidis, I.} \& \textsc{Firth, D.} (2010).
\newblock A generic algorithm for reducing bias in parametric estimation.
\newblock \textit{Electronic Journal of Statistics} \textbf{4}, 1097--1112.

\bibitem[{Lehmann \& Romano(2005)}]{lehm2005}
\textsc{Lehmann, E.~L.} \& \textsc{Romano, J.~P.} (2005).
\newblock \textit{Testing Statistical Hypotheses}.
\newblock Springer.

\bibitem[{McCullagh \& Tibshirani(1990)}]{mccu1990}
\textsc{McCullagh, P.} \& \textsc{Tibshirani, R.} (1990).
\newblock A simple method for the adjustment of profile likelihoods.
\newblock \textit{Journal of the Royal Statistical Society Series B}
  \textbf{52}, 325--344.

\bibitem[{Oja(2013)}]{oja2013}
\textsc{Oja, H.} (2013).
\newblock Multivariate median.
\newblock In \textit{Robustness and Complex Data Structures}, C.~Becker,
  R.~Fried \& S.~Kuhnt, eds. Springer, Berlin, pp. 3--15.

\bibitem[{Pace \& Salvan(1992)}]{pacesalvan1992}
\textsc{Pace, L.} \& \textsc{Salvan, A.} (1992).
\newblock A note on conditional cumulants in canonical exponential families.
\newblock \textit{Scandinavian Journal of Statistics} \textbf{19}, 185--191.

\bibitem[{Pace \& Salvan(1997)}]{pace:salv:1997}
\textsc{Pace, L.} \& \textsc{Salvan, A.} (1997).
\newblock \textit{Principles of Statistical Inference from a Neo-Fisherian
  Perspective}, vol.~4.
\newblock World Scientific Pub Co Inc.

\bibitem[{Pace \& Salvan(1999)}]{pace1999}
\textsc{Pace, L.} \& \textsc{Salvan, A.} (1999).
\newblock Point estimation based on confidence intervals: exponential families.
\newblock \textit{Journal of Statistical Computation and Simulation}
  \textbf{64}, 1--21.

\bibitem[{Read(1985)}]{read1985}
\textsc{Read, C.~B.} (1985).
\newblock Median unbiased estimators.
\newblock In \textit{Encyclopedia of Statistical Sciences}, S.~Kotz, N.~Johnson
  \& C.~Read, eds., vol.~5. Wiley, New York, pp. 424--426.

\bibitem[{Sartori(2003)}]{sart2003}
\textsc{Sartori, N.} (2003).
\newblock Modified profile likelihoods in models with stratum nuisance
  parameters.
\newblock \textit{Biometrika} \textbf{90}, 533--549.

\bibitem[{Sartori(2006)}]{sartori2006}
\textsc{Sartori, N.} (2006).
\newblock Bias prevention of maximum likelihood estimates for scalar skew
  normal and skew t distributions.
\newblock \textit{Journal of Statistical Planning and Inference} \textbf{136},
  4259--4275.

\bibitem[{Stern(1997)}]{stern1997}
\textsc{Stern, S.~E.} (1997).
\newblock A second-order adjustment to the profile likelihood in the case of a
  multidimensional parameter of interest.
\newblock \textit{Journal of the Royal Statistical Society Series B}
  \textbf{59}, 653--665.

\end{thebibliography}

\section*{Supplementary material}
\label{SM}

Supplementary material includes some discussion on the discrete case and details and quantities for the implementation of the method, together with additional simulation results for Examples 4, 7 and 8.

%%%%%%%%%%%%%%%%%%%%%%%%%%%%%%%%%
\section{Comparison with exact median unbiased estimator in  simple binomial regression models}
%%%%%%%%%%%%%%%%%%%%%%%%%%%%%%%%%

Consider first a simple binomial regression model with $y_1,\ldots,y_5$  realizations of independent $Bi(m,\pi_i)$ random variables, with $\log\{\pi_i/(1-\pi_i)\}= \theta \,x_i$, $i=1,\ldots,5$, with covariate values $(x_1,\ldots,x_5)=(-0.560, -0.230,  0.071,  0.129,  1.559)$, generated from a standard normal distribution. The sufficient statistic is $t=\sum_{i=1}^5 y_i x_i$ and takes $(m+1)^5$ distinct values.

We compare the maximum likelihood estimator, $\hat\theta$, which amounts to considering only the leading term of the Cornish-Fisher expansion for the median of $U(\theta)$, and the median bias reduced estimator, $\tilde\theta$, with the exact median unbiased estimator,  $\tilde\theta^e$, %$\tilde\theta^e$, 
for increasing values of $m$. All three estimators vary monotonically with $t$ and the latter estimator \citep[see, for instance,][]{hirji1989}  is defined as  $\tilde\theta^e=(\theta_{*}+\theta_{**})/2$, where $\theta_{*}$ and $\theta_{**}$ are such that
$$
\P_{\theta_{*}}(T \leq t) \geq 1/2, \qquad
\P_{\theta_{**}}(T \geq t) \geq 1/2 \,.
$$
When $t$ is equal to either the maximum or the minimum of its possibile values, then only one of $\theta_{*}$ or $\theta_{**}$ is defined. 
In such case, $\tilde\theta^e$ is taken to be $\theta_{*}$ or $\theta_{**}$, whichever exists. 
This estimator satisfies 
$$
\P_\theta(\tilde\theta^e \leq \theta) \geq 1/2, \qquad
\P_\theta(\tilde\theta^e \geq \theta) \geq 1/2 \,.
$$

%In the following, we compare the maximum likelihood estimator, $\hat\theta$, and the median bias reduced estimator, $\tilde\theta$, with the exact median unbiased estimator, $\hat\theta_u$, for increasing values of $m$. The latter estimator \citep[see, for instance,][]{hirji1989}  is such that
%$$
%\P_\theta(\hat\theta_u \leq \theta) \geq 1/2, \qquad
%\P_\theta(\hat\theta_u \geq \theta) \geq 1/2 \,.
%$$
%
%In the present example all three estimators above vary monotonically with $t$ and for the computation of $\hat\theta_u$ we first find $\theta_{*}$ and $\theta_{**}$ such that
%$$
%\P_{\theta_{*}}(T \leq t) \geq 1/2, \qquad
%\P_{\theta_{**}}(T \geq t) \geq 1/2 \,.
%$$
%and then define $\hat\theta_u=(\theta_{*}+\theta_{**})/2$. When $t$ is equal to either the maximum or the minimum of the possibile values of $T$, then only one of $\theta_{*}$ or $\theta_{**}$ is defined. 
%In such case, $\hat\theta_u$ is taken to be $\theta_{*}$ or $\theta_{**}$, whichever exists. 
%
%
%

\begin{figure}\centering
%\vspace{-1cm}
\includegraphics[width=0.75\textwidth]{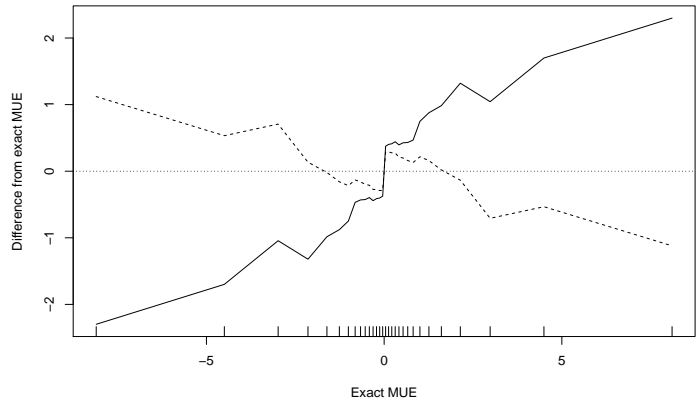}
\includegraphics[width=0.75\textwidth]{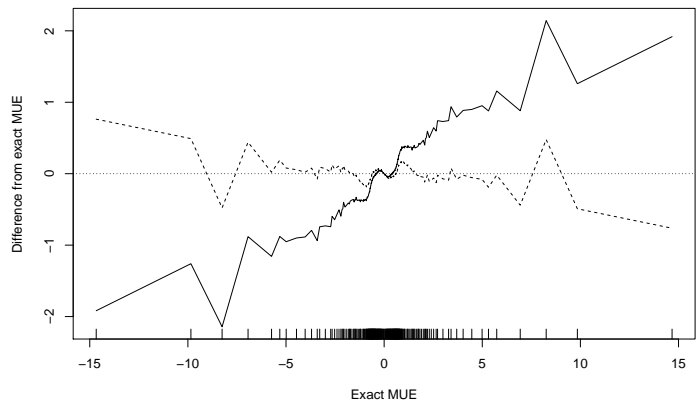} 
\includegraphics[width=0.75\textwidth]{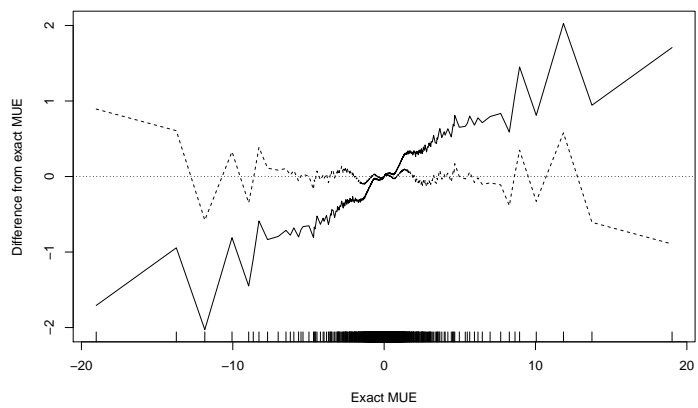} 
%\vspace{-1.cm}
\caption{Simple binomial regression. Differences $\hat\theta-\tilde\theta^e$ (solid) and $\tilde\theta-\tilde\theta^e$ (dashed) as a function of  $\tilde\theta^e$ when $m$ is equal to 1 (top), 2 (middle) and 3 (bottom). Ticks on the horizontal axes represent values of $\tilde\theta^e$.}
\label{fig-osc1}
\end{figure}

For $m=1,2,3$, $t$ takes respectively $32$, $243$ and $1024$ distinct values. Figure \ref{fig-osc1} shows the differences $\hat\theta-\tilde\theta^e$ and $\tilde\theta-\tilde\theta^e$ as functions of $\tilde\theta^e$ in the three situations. We note that the two points corresponding to the minimum and maximum values of $t$ are not reported since $\hat\theta$ is respectively $-\infty$ or $+\infty$. The proposed estimator $\tilde\theta$ is closer to $\tilde\theta^e$ than $\hat\theta$ in all three situations, with relative differences getting smaller as the number of points in the sample space increases.

%{hirji1989} 

As an example with $p>1$, consider the hypothetical clinical trial data in \citet[][Table 2]{hirji1989} with $n=30$ patients belonging to two age groups (age less or equal than 30 years, and age greater than 30 years) of size 20 and 10, respectively. Each group is randomized to receive one of two treatments, with 9 and 6 patients receiving the first treatment in the first and second age group, respectively. Let $y_i$ be the binary disease outcome ($y_i=1$ for a positive outcome, $y_i=0$ otherwise). Moreover, let $x_{i2}$ be a binary age indicator  ($x_{i2}=1$ if age  is less or equal to 30 years, $x_{i2}=0$ if age is  greater than 30 years) and $x_{i3}$ be a binary treatment indicator ($x_{i3}=1$ for the first treatment, $x_{i3}=0$ for the second treatment), $i=1,\ldots,30$. Then, with $\pi_i$ the probability of a positive outcome, a logistic model relating the response of the $i$-th patient to treatment and age can be written as
$$
\log\{\pi_i/(1-\pi_i)\}= \beta_1+\beta_2x_{i2}+\beta_3x_{i3}\,,\qquad i=1,\ldots,30\,.
$$
Here $\beta_3$ is the relative log odds of response for treatment 1 versus treatment 2, and can be considered as the parameter of interest. The exact conditional median unbiased estimator $\tilde\beta_3^e$ \citep{hirji1989} can be obtained using the definition above, applied to the conditional distribution of $t=\sum_{i=1}^{30} x_{i3}Y_i$ given $s=(\sum_{i=1}^{30} y_i, \sum_{i=1}^{30} x_{i2}y_i)$. As in \citet[][Table 2]{hirji1989}, we compare in Table \ref{hirji}   $\hat\beta_3$, $\tilde\beta_3$ and $\tilde\beta_3^e$, for all possible values of $t$ in the conditional distribution of $T$ given $s=(16, 12)$.   Estimate $\tilde\beta_3$ is the third component of the joint bias reduced estimate $\tilde\beta$, as in Example 7, and it is third order equivalent to $\tilde\beta_{3_P}$, from (8). The bias reduced estimator is uniformly closer to the exact conditional median unbiased estimator than the maximum likelihood estimate.

\begin{table}\centering
\def~{\hphantom{0}}
\caption{Values of $T$ given $s=(16, 12)$ and corresponding values of $\hat\beta_3$, $\tilde\beta_3$ and $\tilde\beta_3^e$ }{
\begin{tabular}{cccc}
%\hline
$t$  & $\tilde\beta_3^e$ & $\tilde\beta_3$ & $\hat\beta_3$ \\ 
%\hline
1 & -4.489 & -6.077 & $-\infty$ \\ 
2 & -3.885 & -3.909 & -4.537 \\ 
3 & -2.876 & -2.900 & -3.239 \\ 
4 & -2.141 & -2.150 & -2.361 \\ 
5 & -1.517 & -1.520 & -1.654 \\ 
6 & -0.953 & -0.955 & -1.032 \\ 
7 & -0.421 & -0.421 & -0.453 \\ 
8 & 0.104 & 0.103 & 0.114 \\ 
9 & 0.641 & 0.640 & 0.695 \\ 
10 & 1.220 & 1.217 & 1.325 \\ 
11 & 1.899 & 1.885 & 2.068 \\ 
12 & 2.851 & 2.778 & 3.103 \\ 
13 & 3.430 & 4.966 &  $+\infty$ \\ 
%\hline
\end{tabular}
}
 \label{hirji}
%\begin{tabnote}
 % \end{tabnote}
\end{table}

%%%%%%%%%%%%%%%%%%%%%%%%%%%%%%%%%
\section{Binary regression}
%%%%%%%%%%%%%%%%%%%%%%%%%%%%%%%%%

We give details on the computation of the needed quantities for (8) and (9), used in Examples 4 and 7 of the paper, respectively. We assume $y_i$, $i=1,\ldots,n$, as independent realizations of binary random variables with probability $\pi_i=F(\eta_i)$, where $\eta_i= x_i \beta$, $x_i=(x_{i1},\ldots, x_{ip})$ is a row vector of covariates, $\beta=(\beta_1,\ldots,\beta_p)^\T$ and $F$ is a known cumulative distribution function. Below,   indices $r$, $s$, and $t$ refer to the components of   $\beta$. We have
  \begin{eqnarray*}
U_r &=&\sum_{i=1}^n x_{ir} A(\eta_i)\{ y_i-F(\eta_i)\}\,,\qquad 
i_{rs}= \sum_{i=1}^n x_{ir} x_{is} A(\eta_i) F'(\eta_i)\,,\\
\nu_{rs,t} &=& \sum_{i=1}^n x_{ir} x_{is}x_{it} B(\eta_i) F'(\eta_i)\,,\\
\nu_{r,s,t}&=& \sum_{i=1}^n x_{ir} x_{is}x_{it} A(\eta_i)^3 F(\eta_i)\{ 1-F(\eta_i)\} \{ 1-2F(\eta_i)\} \,,
\end{eqnarray*}
with
\begin{eqnarray*}
A(\eta_i) &=&\frac{F'(\eta_i)}{F(\eta_i)\{ 1-F(\eta_i)\}}\,,\\
B(\eta_i) &=&\frac{F''(\eta_i)}{F(\eta_i)\{ 1-F(\eta_i)\}}+\frac{F'(\eta_i)^2\{ 2F(\eta_i)-1\}}{F(\eta_i)^2\{ 1-F(\eta_i)\}^2} \,,
\end{eqnarray*}

\noindent
where $F'(\cdot)$ and $F''(\cdot)$ are first and second derivatives of $F(\cdot)$. If $F(\cdot)$ is the logistic cumulative distribution function, $A(\cdot)=1$ and $B(\cdot)=0$.

Ingredients of Fisher scoring equation (13) may be written in matrix form as
$U(\beta) = X^\T W(\beta) v(\beta)$ and $i(\beta)=X^\T W(\beta) X$, where $X$ is the design matrix with entries $x_{ir}$, $W(\beta)$ is a diagonal matrix with diagonal elements $\{F'(\eta_i)\}^2/[F(\eta_i)\{ 1-F(\eta_i)\}]$, and $v(\beta)$ is a vector with elements $v_i(\beta)=\{ y_i-F(\eta_i)  \}/ F'(\eta_i)$, $i=1,\ldots,n, r=1,\ldots,p$. Hence, (13) becomes
$$
\tilde \beta^{(k+1)} = \{X^\T W(\tilde\beta^{(k)}) X\}^{-1} X^\T W(\tilde\beta^{(k)}) \tilde v(\tilde\beta^{(k)})\,,
$$
where the adjusted response variable $\tilde v(\tilde\beta^{(k)})= X\{\tilde\beta^{(k)} + M_1(\tilde\beta^{(k)})\} + v(\tilde\beta^{(k)})$ includes the modification term $M_1(\tilde\beta^{(k)})$.% evaluated at the current estimate.
 
Simulation results for Example 7 with probit link are in Table \ref{kenne:tab1}.

\begin{table}\centering\footnotesize
 \begin{threeparttable}
 \def~{\hphantom{0}}
\caption{Simulation results for endometrial cancer study  with probit link.  
For  maximum likelihood,  B, RMSE and coverage are conditional upon finiteness of the estimates}{
\begin{tabular}{lcccccc}
              &             PU &       MAE &         B &     RMSE & Wald  & Score \\                                      
$\hat\beta$  & 43.1 & 0.57 & 0.21 & 0.96 & 95.3 & 95.3 \\ 
                      & 43.4 & 0.38 & 0.44 & 1.55 & 97.1 & 95.0 \\ 
                      & 50.5 & 0.01 & -0.00 & 0.02 & 94.1 & 94.2 \\ 
                      & 58.1 & 0.33 & -0.18 & 0.61 & 95.5 & 95.4 \\                        
$\hat\beta^*$& 52.8 & 0.51 & -0.01 & 0.80 & 95.90 & -- \\ 
                       & 51.7 & 0.33 & 0.01 & 0.52 & 97.10 & -- \\ 
                       & 49.2 & 0.01 & -0.00 & 0.02 & 96.40 & -- \\ 
                      & 45.0 & 0.30 & 0.01 & 0.48 & 94.50 & -- \\                            
$\tilde\beta$&50.5 & 0.53 & 0.05 & 0.85 & 96.1 & 95.2 \\ 
                      & 49.3 & 0.34 & 0.06 & 0.58 & 97.0 & 94.9 \\ 
                     & 50.1 & 0.01 & -0.00 & 0.02 & 96.0 & 94.2 \\ 
                     & 49.8 & 0.31 & -0.06 & 0.51 & 95.6 & 95.0 \\\end{tabular}}
\label{kenne:tab1}
%\begin{tablenotes}
%      \small
%      \item For the mle, 85 samples  without convergence
%      \item For the  mean bias correction, 65 samples
%      \item  For the median bias correction 82 samples
%    \end{tablenotes}
\begin{tablenotes}\footnotesize\item
PU, percentage of underestimation; MAE,   median absolute error; B, bias; RMSE, root mean squared error; Wald and Score, percentage coverage of 95\% Wald-type and Score-type confidence intervals.  
\end{tablenotes}
\end{threeparttable}
\end{table}

\section{Beta regression}

We give quantities for computing (9) in Example 8 of the paper.   Let $y_i$, $i=1,\ldots,n$, be independent realizations of beta random variables with parameters $\phi\mu_i$ and $\phi(1-\mu_i)$, i.e. with expected value $\mu_i$ and precision parameter $\phi$. We assume a  regression structure for the expected value $\mu_i=g^{-1}(\eta_i)$, 
%and precision parameter $\phi$, 
where $\eta_i= x_i \beta$, $x_i=(x_{i1},\ldots, x_{ip})$ is a row vector of covariates and $g(\cdot)$ is a given link function, such as the logit. We denote by $\theta=(\beta_1,\ldots,\beta_p,\phi)^\T$ the full vector of parameters.  
%The model density is given by
%$$
%p_{_{Y_i}}(y_i; \mu_i,\phi)=\frac{\Gamma(\phi)}{\Gamma(\mu_i\phi)\Gamma\{(1-\mu_i)\phi\}}y_i^{\mu_i\phi-1}(1-y_i)^{(1-\mu_i)\phi-1}\,,
%$$ with $0<y_i<1$, $0<\mu_i<1$ and $\phi>0.$ 
The log-likelihood has the form
$$
\ell(\theta)=\sum_{i=1}^n\{\mu_i\phi t_i+(1-\mu_i)\phi z_i+\log\Gamma(\phi)-\log\Gamma(\mu_i\phi)-\log\Gamma\{(1-\mu_i)\phi\}\},
$$ where $t_i=\log (y_i)$ and 
$z_i=\log(1-y_i)$.
% are components of  the sufficient statistics with natural parameters $\delta_i^1=\mu_i\phi$ and $\delta_i^2=(1-\mu_i)\phi$. 
 
Let $\bar t_i=t_i- \Psi^{(0)}(\phi\mu_i)+\Psi^{(0)}(\phi)$ and $\bar z_i=z_i-\Psi^{(0)}(\phi(1-\mu_i))+\Psi^{(0)}(\phi)$, 
%with $E_{\theta}(T_i)=\Psi^{(0)}(\phi\mu_i)-\Psi^{(0)}(\phi)$ and $E_{\theta}(Z_i)=\Psi^{(0)}(\phi(1-\mu_i))-\Psi^{(0)}(\phi)$ . 
with $\Psi^{(r)}(k)=d^{r+1}\log\Gamma(k)/dk^{r+1} $ the polygamma function of order $r$,  $r=0,1,2,\ldots$. The needed quantities for (9) are
\begin{equation*}
\begin{split}
&U_{\beta_r} =\phi\sum_{i=1}^nx_{ir}(\bar t_i-\bar z_i) (d\mu_i/d\eta_i),\,\, %r=1,\dots,p, \mbox{  and  } 
\quad U_\phi=\sum_{i=1}^n \mu_i(\bar t_i-\bar z_i)+\bar z_i,\\
%\,\,r=p+1\\ 
&\dot\imath_{\beta_r\beta_s}=\phi^2\sum_{i=1}^n x_{ir}x_{is}[\Psi^{(1)}(\phi\mu_i)-\Psi^{(1)}\{\phi(1-\mu_i)\}](d\mu_i/d\eta_i)^2,\\%\,\,r,s=1,\ldots,p\\
&\dot\imath_{\beta_r\phi}=\phi\sum_{i=1}^nx_{ir}(\mu_i [ \Psi^{(1)}(\phi\mu_i)+\Psi^{(1)}\{\phi(1-\mu_i)\}]-\Psi^{(1)}\{\phi(1-\mu_i)\})(d\mu_i/d\eta_i),\\%\,\,r=1,\ldots,p\\
&\dot\imath_{\phi\phi}=\sum_{i=1}^n[\Psi^{(1)}(\phi\mu_i)\mu_i^2+\Psi^{(1)}\{\phi(1-\mu_i)\}(1-\mu_i)^2]-n\Psi^{(1)}(\phi),\\
%\end{split}
%\end{equation*}
%\begin{equation*}
%\begin{split}
&\nu_{\beta_r,\beta_s,\beta_t}=\phi^3\sum_{i=1}^nx_{ir}x_{is}x_{it}[\Psi^{(2)}(\phi\mu_i)-\Psi^{(2)}\{\phi(1-\mu_i)\}](d\mu_i/d\eta_i)^3,\\%\,\,r,s,t=1,\ldots,p\\
&\nu_{\beta_r,\beta_s,\phi}=\phi^2\sum_{i=1}^nx_{ir}x_{is}[\Psi^{(2)}(\phi\mu_i)\mu_i+\Psi^{(2)}\{\phi(1-\mu_i)\}(1-\mu_i)](d\mu_i/d\eta_i)^2,\\%\,\,r,s=1,\ldots,p\\
&\nu_{\beta_r,\phi,\phi}=\phi\sum_{i=1}^nx_{ir}[\Psi^{(2)}(\phi\mu_i)\mu_i^2-\Psi^{(2)}\{\phi(1-\mu_i)\}(\mu_i-1)^2](d\mu_i/d\eta_i),\\%\,\,r=1,\ldots,p\\
&\nu_{\phi,\phi,\phi}=\sum_{i=1}^n[\Psi^{(2)}(\phi\mu_i)\mu_i^3-\Psi^{(2)}\{\phi(1-\mu_i)\}(\mu_i-1)^3]-n\Psi^{(2)}(\phi),%\\
\end{split}
\end{equation*}
\begin{equation*}
\begin{split}
&\nu_{\beta_r,\beta_s\beta_t}=\phi^2\sum_{i=1}^nx_{ir}x_{is}x_{it}[\Psi^{(1)}(\phi\mu_i)-\Psi^{(1)}\{\phi(1-\mu_i)\}](d\mu_i/d\eta_i)(d^2\mu_i/d\eta_i^2),\\%\,\,r,s,t=1,\ldots,p\\
&\nu_{\beta_r,\beta_s\phi}=\phi\sum_{i=1}^nx_{ir}x_{is}[\Psi^{(1)}(\phi\mu_i)+\Psi^{(1)}\{\phi(1-\mu_i)\}](d\mu_i/d\eta_i)^2,\\%\,\,r,s=1,\ldots,p\\
&\nu_{\beta_r,\phi\phi}=0\\
&\nu_{\phi,\beta_r\beta_s}=\phi\sum_{i=1}^nx_{ir}x_{is}(\mu_i[\Psi^{(1)}(\phi\mu_i)+\Psi^{(1)}\{\phi(1-\mu_i)\}]-\Psi^{(1)}\{\phi(1-\mu_i)\})(d^2\mu_i/d\eta_i^2),\\%\,\,r,s=1,\ldots,p\\
&\nu_{\phi,\beta_r\phi}=\sum_{i=1}^nx_{ir}(\mu_i[\Psi^{(1)}(\phi\mu_i)+\Psi^{(1)}\{\phi(1-\mu_i)\}]-\Psi^{(1)}\{\phi(1-\mu_i)\})(d\mu_i/d\eta_i),\\%\,\,r=1,\ldots,p\\
&\nu_{\phi,\phi\phi}=0.\\
&
\end{split}
\end{equation*}
%where indices $r$, $s$ and $t$ refer to the components of $\beta$.
An {\tt R} implementation of the method, using formula (13) of the paper, is given in function {\tt mbrbetareg}  available
at \texttt{https://github.com/eulogepagui/mbrbetareg}.\\

Complete simulation results for Example 8  are in Table \ref{kenne:tabfood2}.\\

\begin{table}\centering
 \begin{threeparttable}
 \def~{\hphantom{0}}
\caption{Simulation results for food expenditure}{
\begin{tabular}{lcccccc}
 & PU & MAE & B & RMSE & Wald  & Score \\ 
 $\hat\theta$  &  50.0 & 0.15 & 0.00 & 0.22 & 93.1 & 94.3\\
  & 50.6  & 0.00 & 0.00 & 0.00 & 93.5 & 94.8\\
 &  50.5 &  0.02 & 0.00  & 0.04 & 93.2 & 94.5\\
 &   32.7 & 6.25 & 5.46 & 11.79 & 95.1 & 95.1\\
$\hat\theta^*$ & 49.8 & 0.15 & 0.00 & 0.22 & 94.7 & --\\%96.2\\
& 50.2 & 0.00 & 0.00 & 0.00 & 95.2 & --\\%96.3\\
& 50.9 & 0.02 & 0.00 & 0.04 & 94.9 & --\\% 96.2\\
& 56.5 & 5.74  & 0.06  & 9.07 & 91.8 & --\\%95.1\\
$\tilde\theta$& 49.9 & 0.15 & 0.00 & 0.22 & 94.3 & 94.4\\
& 50.3 & 0.00 & 0.00 & 0.00 & 94.8 & 94.8\\
&  50.8 & 0.02 & 0.00 & 0.04 & 94.5 & 94.5\\
&  49.8 & 5.69 & 1.49 & 9.56 & 93.7 & 95.8\\
\end{tabular}}
\label{kenne:tabfood2}
\begin{tablenotes}\footnotesize\item
    PU, percentage of underestimation; MAE, median absolute error; B, bias; RMSE, root mean squared error; Wald and Score, percentage coverage of 95\% Wald-type and score-type confidence intervals.   
      \end{tablenotes}
\end{threeparttable}
\end{table}
As a further example, we consider the gasoline yield data as in \citet[][Section 4.3]{kosmi2010}.   
Here $n=32$ and the response variable is the proportion of crude oil converted to gasoline after
distillation and fractionation. Covariates are  9 indicators representing the 10 distinct experimental settings in the data  and the temperature in degrees Fahrenheit at which all gasoline has vaporized.
Estimates of $\theta=(\beta_1,\ldots,\beta_{11},\phi)^\T$, where $\beta_1$ is an intercept, $\beta_2,\ldots,\beta_{10}$ are the coefficients of the 9 indicators, $\beta_{11}$ is the coefficient of the temperature, and with the logit link are given in Table \ref{tab3}. Values for the regression coefficients  are essentially the same for all methods, while notable differences are observed for the dispersion parameter. These in turn influence estimates of standard errors for the regression coefficients.

We performed a simulation study with the same sample size and covariates as in the gasoline yield data and with parameter fixed at $\hat\theta$.  Results obtained from 100,000 simulated samples in Table \ref{tab4} show marked differences among estimators of $\phi$, with $\tilde\phi$ achieving median centering. These differences also imply different coverages of confidence intervals for the  regression coefficients. In this rather extreme case score-type confidence intervals are numerically very unstable and only results for Wald-type intervals are reported. 
\begin{table}[ht]
\def~{\hphantom{0}}
\caption{Gasoline data. Regression estimates (s.e.).}{
\begin{tabular}{lcccc}
  %\toprule
 & $\hat\theta$ & $\hat\theta^*$  & $\tilde\theta$  \\ 
  %\midrule
$\beta_1$ & -6.160 (0.182) & -6.142 (0.236) & -6.144 (0.228) \\ 
  $\beta_2$ & 1.728 (0.101) & 1.723 (0.131) & 1.724 (0.127) \\ 
  $\beta_3$ & 1.323 (0.118) & 1.319 (0.153) & 1.319 (0.148) \\ 
  $\beta_4$ & 1.572 (0.116) & 1.567 (0.150) & 1.568 (0.145) \\ 
  $\beta_5$ & 1.060 (0.102) & 1.057 (0.132) & 1.058 (0.128) \\ 
  $\beta_6$ & 1.134 (0.104) & 1.130 (0.134) & 1.131 (0.130) \\ 
  $\beta_7$ & 1.040 (0.106) & 1.037 (0.137) & 1.038 (0.133) \\ 
  $\beta_8$ & 0.544 (0.109) & 0.542 (0.141) & 0.543 (0.137) \\ 
  $\beta_9$ & 0.496 (0.109) & 0.494 (0.141) & 0.495 (0.136) \\ 
  $\beta_{10}$ & 0.386 (0.119) & 0.385 (0.154) & 0.385 (0.148) \\ 
  $\beta_{11}$ & 0.011  (0.000)& 0.011 (0.001) & 0.011 (0.001) \\ 
  $\phi$ & 440.278 (110.026) & 261.038 (65.216) & 279.409 (69.809) \\   
  %\bottomrule
\end{tabular}}
\label{tab3}
\end{table}

\begin{table}[ht]\centering
 \begin{threeparttable}
 \def~{\hphantom{0}}
\caption{Simulation results for gasoline data}{
\begin{tabular}{rrrrrr}
 % \toprule
 & PU & MAE & B & RMSE & Wald   \\ 
  %\midrule
 $\hat\theta$&52.63  & 0.124  &-0.015  & 0.183 &86.82\\
 &48.67 &  0.068  & 0.004 &  0.102& 86.86\\
 & 48.14 &  0.081 &  0.005 &  0.119& 86.95\\
 &48.75  & 0.078   &0.004  & 0.116& 87.19\\
 &49.18  & 0.070 &  0.002 &  0.103 &87.12\\
 &49.55 &  0.070  & 0.002  & 0.103& 87.70\\
 & 49.10 &  0.071 &  0.003 &  0.107 &87.20\\
 &49.84  & 0.073  & 0.001  & 0.109& 87.29\\
 & 49.69 &  0.074 &  0.002 &  0.110 &86.57\\
&49.60  & 0.081  & 0.002 &  0.118& 87.23\\
&47.51  & 0.000  & 0.000  & 0.000& 87.14\\
&5.65& 254.489& 302.277& 395.193 &74.92\\
   %\midrule
$\hat\theta^*$& 50.00  & 0.125  &-0.002  & 0.182& 94.76\\
&49.95  & 0.069  & 0.001  & 0.102 &94.69\\
& 49.16&   0.081&   0.002 &  0.119& 94.32\\
&49.81  & 0.078  & 0.000 &  0.116& 94.86\\
& 50.06 &  0.069 &  0.000 &  0.102 &94.73\\
& 50.44  & 0.070 &  0.000  & 0.103& 94.95\\
& 49.82 &  0.071  & 0.001  & 0.106& 94.42\\
& 50.02  & 0.073  & 0.000 &  0.109& 94.45\\
& 49.96 &  0.074 &  0.001 &  0.110& 94.57\\
&49.81  & 0.081  & 0.001  & 0.118 &94.97\\
&49.82  & 0.000 &  0.000&   0.000 &94.67\\
&58.12 & 93.669  & 0.209& 151.152 &84.83\\
   %\midrule
$\tilde\theta$ & 50.30 &  0.125 & -0.004  & 0.182& 93.83\\
&49.67 &  0.069  & 0.002  & 0.102 &93.97\\
&49.08&   0.081&   0.002&   0.119 &93.60\\
& 49.67  & 0.078  & 0.001 &  0.116& 94.04\\
& 49.83 &  0.069 &  0.001 &  0.102& 93.98\\
&50.25 &  0.070  & 0.000 &  0.103& 94.24\\
& 49.64 &  0.071 &  0.001 &  0.106& 93.67\\
&49.95 &  0.073  & 0.001 &  0.109& 93.68\\
&49.88 &  0.074  & 0.001  & 0.110 &93.90\\
&49.81 &  0.081 &  0.001  & 0.118& 94.29\\
&49.51 &  0.000 &  0.000 &  0.000& 94.02\\
&49.76 & 92.138 & 31.176& 164.736& 88.72\\
   %\bottomrule
\end{tabular}}
\label{tab4}
\begin{tablenotes}\footnotesize\item
 PU, percentage of underestimation; MAE, median absolute error; B, bias; RMSE, root mean squared error; Wald, percentage coverage of 95\% Wald-type confidence intervals.   
\end{tablenotes}
\end{threeparttable}
\end{table}

\end{document}